\documentclass[printer]{aa}
\usepackage[dvips]{graphicx}
\usepackage{txfonts}
\usepackage{natbib}
\usepackage{longtable}
\bibpunct{(}{)}{;}{a}{}{,}
\begin{document}

\title{Circular polarization survey of intermediate polars I. Northern targets
  in the range 17h$<$R.A.$<$23h
\thanks{Based on observations obtained at the Nordic
Optical Telescope at the Roque de los Muchachos Observatory in La Palma.}
}

\titlerunning{Circular polarization survey of intermediate polars I.}

\author{
  O.W.~Butters\inst{1,2}
\and
  S.~Katajainen\inst{3}
\and
  A.J.~Norton\inst{1}
\and
  H.J.~Lehto\inst{3}
\and
  V.~Piirola\inst{3}}

\authorrunning{Butters et al.}

\institute{
  Department of Physics and Astronomy, The Open University, Walton Hall,
  Milton Keynes MK7 6AA, UK
\and
Department of Physics and Astronomy, University of Leicester,
  Leicester, LE1 7RH, UK\\
\email{oliver.butters@star.le.ac.uk}
\and
  Tuorla Observatory, Department of Physics and Astronomy, University
  of Turku, V\"ais\"al\"antie 20, FI-21500 Piikki\"o, Finland\\
  \email{sekataja@utu.fi}}

\date{Accepted 2007 ???;
      Received  2007 ???;
      in original form 2007 ???}

\abstract
%context
{The origin, evolution, and ultimate fate of magnetic cataclysmic
  variables are poorly understood. It is largely the nature of the magnetic
  fields in these systems that leads to this poor
  understanding. Fundamental properties, such as the field strength and
  the axis alignment, are unknown in a majority of these systems.}
%aims
{We undertake to put all the previous circular polarization
  measurements into context and systematically survey intermediate polars for signs of circular
  polarization, hence to get an indication of their true magnetic field
  strengths and try to understand the evolution of magnetic
  cataclysmic variables.}
%method
{We used the TurPol instrument at the Nordic Optical Telescope to
  obtain simultaneous UBVRI photo-polarimetric observations of a set of intermediate
  polars, during the epoch 2006 July 31 - August 2.}
%results
{Of this set of eight systems two (\object{1RXS~J213344.1+510725} and
  \object{1RXS~J173021.5--055933}) were found to
  show significant levels of circular polarization, varying with spin phase. 
  Five others (\object{V2306~Cyg},
  \object{AO~Psc}, \object{DQ~Her}, \object{FO~Aqr}, and
  \object{V1223~Sgr}) show some evidence for circular polarization
  and variation of this with spin phase, whilst \object{AE~Aqr}
  shows little evidence for polarized emission. We
  also report the first simultaneous UBVRI photometry
  of the newly identified intermediate polar 1RXS~J173021.5--055933.}
%conclusions
{Circular polarization may be ubiquitous in intermediate polars, albeit at
a low level of one or two percent or less. It is stronger at longer
wavelengths in the visible spectrum. Our results lend further support to
the possible link between the presence of soft X-ray components and the
detectability of circular polarization in intermediate polars.}


\maketitle

\section{Introduction to magnetic cataclysmic variables}
Cataclysmic variables (CVs) are a class of binary system in which a main sequence
star (the secondary) transfers matter to its white dwarf (WD) companion (the
primary), via Roche lobe overflow. Within
this class there is a division of systems based upon the magnetic
field strength of the WD, the non-magnetic (or only very slightly magnetic) CVs
and the magnetic systems (mCVs). This is an important distinction, as
the presence of a magnetic field on the WD has a profound effect on the structure and evolution
of the system, since the accreting material is a plasma, its motion is
subject to the field. In mCVs the
accreting material typically attaches itself to the magnetic field lines 
close to the magnetospheric radius and is
funnelled toward the WD surface. If the WD has a dipole magnetic field
structure then this has the effect
of concentrating the accreting material into small regions above the magnetic
poles. The material reaches supersonic speeds as it is accelerated
toward the surface under the influence of gravity. It then goes
through a shock as it decelerates to subsonic speeds. This matter then
cools as it falls to the surface, emitting hard X-rays
($kT\sim10-60\rm{keV}$) via bremsstrahlung radiation and
optical/infrared cyclotron radiation. Many mCVs also exhibit a soft
X-ray component, which is thought to occur due to either some of the accreting
material somehow avoiding the shock and directly impacting the surface
of the WD (see e.g. \citet{motch96}), or from the reprocessing of hard
X-rays (see e.g. \citet{lamb79}). Recently \citet{evans07} have
suggested that this component is in fact present in all mCVs, and that its detection 
depends on whether geometric factors allow it to be seen or not.

The mCVs are themselves divided into two groups based upon their
periodicities, which are in turn assumed to imply different
magnetic field strengths. The polars contain synchronously
rotating WDs, whilst in intermediate polars (IPs) the WDs rotate asynchronously
with respect to the binary orbit.

The polars typically have a magnetic field strength of the order
10--100 MG. This has the effect of channelling the accreting material onto the WD surface before an accretion disc can form. Furthermore, the magnetic field of the WD extends to
the body of the secondary and will act to synchronize the spin and
orbital periods of the system. Typically the polars are found to have
an orbital and spin period of less than two hours. For an overview of
polars see e.g. \citet{cropper90}.

IPs are asynchronous systems, generally
{\it believed} to have a magnetic field strength of approximately 1--10 MG. This smaller field means that matter falls inward toward the WD in much the same
way as in non-magnetic CVs. However, at some critical distance from the WD, the magnetic
field will begin to dominate, and the accretion process is
altered. It then carries on in much the same way as in the polars. For
a comprehensive review of IPs see e.g. \citet{warner95} or
\citet{patterson94}.

Non-magnetic CVs evolve from long orbital periods (a few hours) to
short orbital periods (just over an hour). The proposed explanation for this
evolution is magnetic braking and gravitational radiation \citep{king88}. Both these
processes should also be present in mCVs and since IPs generally have a
longer orbital period than the polars it was
suggested that IPs evolve into polars \citep{chanmugam84}. This hypothesis
has not been widely accepted as there is little evidence of strong
magnetic fields in IPs. There are however some IPs with a similar magnetic field strength to the polars, up to 30~MG in the case
of V405~Aur \citep{piirola08}.

\citet{cumming04} suggested that the relatively high accretion rate in IPs may
suppress the magnetic field by overcoming ohmic diffusion, such that
magnetic flux is advected into the interior of the WD. This would
cause the surface magnetic field to appear less than it really
is. Considering the evolution of non-magnetic CVs, i.e. the orbital
period decreases until $\sim3$~hr when the accretion
effectively `turns off' then it resumes at $\sim2$~hr, allows a
regime where an IP may have its `true' magnetic field resurface
when the accretion turns off, allow synchronization, then reappear as
a polar when the accretion resumes at a lower rate. This theory also
ties in with studies of isolated WDs, where peak magnetic fields of
$\sim10^9$~G are seen, whereas in binaries the field strength is
typically an order of magnitude smaller (the maximum seen is 230~MG in
\object{AR~UMa} \citep{schmidt96}) \citep{wickramasinghe00}.

In order to test any evolutionary hypotheses an accurate
determination of the magnetic fields present in IPs is
needed. The best way of doing this is via circular polarization
measurements.

Here we report the first paper in our survey of circular polarization
emission from IPs. Initially we summarise the field as it stands, in
order to put our survey in context (Sec.~\ref{background}). Then we outline the method we use
and the style of reporting our data (Sec.~\ref{observations}). Then we report the results of
each of the targets (Sec.~\ref{results}),
followed by a discussion of each (Sec.~\ref{discussion}). We have further data
in hand for some of the Southern Hemisphere targets that we will be
reporting in the near future.

\section{Circular polarization}
\label{background}

\subsection{mCVs}

In an accretion column in a mCV there will be material accreting onto
the magnetic poles of the WD. The electrons in this material will
spiral along the magnetic field lines as they fall to the surface,
emitting circularly polarized radiation. The
inhomogeneity of the magnetic field structure, the velocity of the
electrons, and their varying
temperature will have the effect of causing the
electrons to emit a broad harmonic structure. By modelling the
polarized radiation given off (i.e. simulating the harmonic structure)
it is possible to estimate the magnetic field strength of the
WD. \citet{Wickramasinghe85} did this for a range of magnetic field
strengths in polars. Their models take into account the magnetic field
strength, the temperature, the plasma parameter, and the viewing
angle. These four variables cause the modelling to give non-unique solutions for the
fitting of the harmonic curve. Their model has been used
extensively in estimating the magnetic field strength in polars by
measuring the level of circular polarization in different wave-bands
(see e.g. \citet{piirola87a, piirola87b, katajainen03}) and has also
been used in IPs (see e.g. \citet{piirola93}).

The polarized light measured in mCVs is generally quoted as a fraction of the total
incoming radiation. In polars the source of
the unpolarized and polarized radiation should vary at the same
  period - the orbital period. In IPs however, the situation is more complex. The presence of
an accretion disc, that emits at optical frequencies, will dilute the
measured polarization. If the flux from the accretion disc varies at any period
other than the spin period then this will dilute the signal in a
complex fashion. Added to this are the possible presence of a hot spot and
the emission from the secondary which will vary at the orbital period.

The geometry of
the accretion column is also an issue, since the circular polarization
is not given off in all directions (see e.g. \citet{norton02}).
 The specific
size/shape of the accretion column will therefore affect the
emission.

Another likely complexity is that the magnetic field structure of the 
WD may not be dipole-like. This could lead to multiple accretion columns at
multiple magnetic poles. If the magnetic field is close to being a
dipole it is also possible that it could be offset from the centre,
leading to one pole appearing to be stronger than the other and not at
diametrically opposed poles. This complexity is further confounded by the very fact that the accreting material does not
  come from infinity, in which case even an ideal dipole would not
  form an accretion column at exactly the position of the magnetic poles.

Given all these complications we cannot be certain what the average magnetic field strength
  {\it really} is in IPs, it is possible that estimates of the field strength
are over an order of magnitude understated.
What we are then assuming to be low
number harmonics of a low magnetic field may be high number harmonics
of a much larger field.

\subsection{Method of circular polarization detections}

Assuming circular polarization in IPs can be measured in a similar way
to polars and that IPs are of a comparable field strength, implies that
harmonics will be present in the UV-IR. The optimal method to reveal
these harmonics would be spin phase resolved circular spectro-polarimetry. This
requires very large telescopes with very specialised instrumentation
allowing high time resolution data collection. This is beyond the
scope of this study, therefore we
concentrate on the more readily available technique of circular photo-polarimetry.

The general principle of circular photo-polarimetry is to measure the
fraction of polarized radiation after a $\lambda /4$ wave plate, which
converts circular polarization into linear polarization. This process introduces biases however, for example, incomplete
90$^\circ$ retardation by the wave plate. These biases may
cause so called `Stokes parameters cross talks', particularly
in the case of targets with non-negligible linear
polarization. These cross talks can effectively be
eliminated by rotating the wave plate to at least two different wave plate
angles and then calculating a single measurement of the circular
polarization by using flux values from both positions. This process has an important effect on the temporal resolution of the data, since enough
time must be spent integrating on the target to get a good signal to
noise value, and if this has to be done multiple times then the
temporal resolution will suffer.

By taking simultaneous circular polarization measurements in different pass-bands a clearer understanding of the harmonic
  structure may be gained. If the pass-bands are defined to have a
  range comparable to the expected width of the harmonics then
  comparison of the circular polarization in each band gives an
  indication of the magnetic field strength.

The method of reporting the level of circular polarization is rather
ambiguous, some authors quote the average level throughout their
observing run (e.g. \citet{stockman92}), some give orbitally binned
data, but most report data phase binned at the spin period.
 Each method may have the effect of
giving a different interpretation of the magnetic field (see
e.g. \citet{uslenghi01} where the data is presented in mean, spin
binned and orbitally binned format). The average value approach could
potentially smooth a
sinusoidal-like variation, with an arbitrary amplitude and zero offset, to an average of
zero. Phase binning at the orbital period may be severely affected by the
variation in orbitally varying unpolarized light. Phase
binning over the spin period can reduce both these effects, and since
the circular polarization is thought to originate from the accretion
column (emission from which varies at the spin period) this is therefore the most desirable approach.
The base line chosen needs to be sympathetic to the orbital variation,
i.e. either be short in relation to the orbital period, or considered
in chunks, and the integration time short compared to the spin period.

\subsection{Previous Circular polarization detections in IPs}

\object{DQ~Her} was the first IP (although not
classified as an IP at the time) to have circular
polarization detected \citep{swedlund74}. This white
light detection was carried out over the course of three months
allowing many measurements over the entire 4.6~hr orbital period. The
level of circular polarization was found to vary periodically over the
spin period (142~s) and to be both positive and negative. This
variation was found to have a different profile over the orbital period
also. The maximum amplitude of variation was found to be approximately
0.6\%. \citet{stockman92} also measured the level of circular
polarization on \object{DQ~Her}, they found a mean level of $+0.01\pm0.01\%$,
however, they remark that short period systems will have their levels
of reported circular polarization reduced due to the long measurement
times.

The next detection of circular polarization in an IP was \object{BG~CMi} \citep{penning86, west87}. Measurements were taken in five
different pass bands at various times over four months. This allowed the
measurements to be plotted over the orbital period (3.75~hr). In the
$1.10-1.38\mu$m band the data were distributed randomly about the mean
of $-1.74\pm0.26\%$ indicating no orbital modulation. Phase binning the data at the spin period
(15.9~min) showed a coherent modulation, however the variation was
within two sigma of being zero. \citet{west87} also considered the
variation of the circular polarization with the passband, they found
that the amplitude detected increased with wavelength, ranging from
$-0.053\pm0.051\%$ at $0.32-0.86\mu$m to $-4.24\pm1.78\%$ at
$1.40-1.65\mu$m.

\object{PQ~Gem} (RE~0751+14) was found in the early 1990's
to exhibit significant circular polarization which was modulated at the spin
period \citep{rosen93, piirola93, potter97}.

Since then, several IPs have been found to exhibit similar behaviour, or have had upper
limits placed on their circular polarization (see Table~\ref{already_known_CP_IPs} for a
summary of all measurements)\addtocounter{table}{1}.

\object{V2306~Cyg} is the only reported IP to show significant
positive polarization in one band and negative in another
\citep{norton02}, this requires two opposite poles to be seen and for
them to be in different states (i.e. one or more of the temperature,
geometry, accretion rate etc must be different at the two poles). In all
the other cases the polarization has the same sign across the
different wave bands.

Each significant (non-zero) measurement of circular polarization has had an estimate of the
magnetic field present on the WD attributed to it. Different
approaches to modelling the emission have lead to differing inferred
values however (see Table~\ref{mag_field_strengths}).

\begin{table}
  \centering
    \caption{Inferred magnetic field strengths from the measured
    circular polarization in IPs.}
    \label{mag_field_strengths}
    \begin{tabular}{lcl}
      \hline\hline
      Name      & Inferred magnetic   & Reference\\
                & field strength (MG) & \\
      \hline
      BG~CMi    & 5--10               & \cite{west87}\\
                & 3--10               & \cite{chanmugam90}\\
      PQ~Gem    & 8--18               & \cite{piirola93}\\
                & 9--21               & \cite{vaeth96}\\
                & 9--21               & \cite{potter97}\\
      V2400~Oph & $\>8$               & \cite{buckley95}\\
                & 9--27               & \cite{vaeth97}\\
      V405~Aur  & $\sim 30$           & \cite{piirola08}\\
      \hline
%      \multicolumn{3}{l}{References  - (1) \cite{west87}; (2) \cite{chanmugam90};}\\
%      \multicolumn{3}{l}{(3) \cite{piirola93}; (4) \cite{vaeth96};}\\
%      \multicolumn{3}{l}{(5) \cite{potter97}; (6) \cite{buckley95}; (7) \cite{vaeth97};}\\
%      \multicolumn{3}{l}{(8) \cite{piirola08}.}\\
    \end{tabular}
\end{table}

Given the rather disparate nature of previous studies, which generally
tend to have information lost in the style of reporting, we have initiated a survey to conclusively measure
the degree of circular polarization and comprehensively characterise its nature in IPs.

\section{Observations}
\label{observations}

Observations were carried out at the 2.56m Nordic Optical Telescope (NOT) on
the island of La Palma over three consecutive nights starting 2006
July 31. The telescope was fitted with the TurPol instrument. This is the double image chopping
polarimeter
\citep{piirola73, piirola88, korhonen84}, which is able to perform
simultaneous photo-polarimetric measurements in all {\sl UBVRI} bands,
by using four dichroic filters (which split the light into five spectral
pass-bands). The pass-bands are defined as having an effective
wavelength of 360, 440, 530, 690, and 830 nm for each of {\sl UBVRI}
respectively. By inserting a plane parallel calcite plate into the beam before the focal
plane, polarization measurements are possible.
The calcite splits the
incoming light into two components, the ordinary and the extra-ordinary,
which are orthogonally linearly polarized.
A diaphragm in the instrument has two apertures,
one passes the star's ordinary component, the other passes the
extra-ordinary component.
A chopper opens and closes the
apertures alternately, illuminating the photo-cathode of the photomultiplier tube.
 By measuring the relative intensities of components after a wave-plate,
(which may be rotated in 90$^\circ$ steps) the degree of circular polarization of the light entering from the star can be calculated.
Both components of the sky background pass both diaphragms, and the polarization of the sky
is thus directly eliminated. In addition, measurements of empty sky
are also done at 10 minute intervals, as this sky value is needed in calibration of the photometry.

For one polarization data point, normally at least four multiples of the
integration time plus mechanical dead-time (few seconds) generated from rotation of the
wave plate is needed, assuming that the circular polarization is measured from two different wave plate positions.
This will have the effect of `smearing' out the
data on some very short period objects, and therefore may under
report their true polarization value. In this study, as some of the targets show remarkable variability
within a short timescale, we have chosen in those cases to use only one wave
plate position measurement, instead of two. The
reduction software was altered in such a way that it could take a
circular polarization measurement from only one position of the
wave plate. This single position polarization measurement improves the temporal resolution of the data, but at
the cost of increasing the uncertainty on the measurement as the biases are
not cancelled out.

The targets chosen for this northern hemisphere survey are those in
Table~\ref{target_list}, and the observing log in
Table~\ref{observing_log}. Note the very short period systems (AE Aqr
(33~s), DQ Her (142~s), and RX1730 (128~s)) have one orientation of
the wave-plate per measurement, and the other targets have two.

\begin{table}
  \centering
    \caption{Target list.}
    \label{target_list}
    \begin{tabular}{llllr@{.}lr@{.}l}
      \hline\hline
      Name               & $\alpha$2000 & $\delta$2000 & V       & \multicolumn{2}{l}{$\rm P_{spin}$} & \multicolumn{2}{l}{$\rm P_{orb}$}\\
	               &              &              & (mag)   & \multicolumn{2}{l}{(s)}            & \multicolumn{2}{l}{(h)}\\
      \hline
      RXJ1730            & 17:30:21     & $-$05:59:32  & 15.8    &  128&0			  & 15&42\\
      \object{DQ Her}    & 18:07:30     & $+$45:51:32  & 13      &  142&1			  &  4&65\\
      \object{V1223 Sgr} & 18:55:02     & $-$31:09:48  & 13.2    &  745&6			  &  3&37\\
      \object{V2306 Cyg} & 19:58:14     & $+$32:32:42  & 16      & 1466&7			  &  4&35\\
      \object{AE Aqr}    & 20:40:09     & $-$00:52:16  & 12      &   33&1			  &  9&88\\
      RXJ2133            & 21:33:44     & $+$51:07:24  & 15.3    &  570&8			  &  7&19\\
      \object{FO Aqr}    & 22:15:55     & $-$08:21:05  & 13.5    & 1254&5			  &  4&85\\
      \object{AO Psc}    & 22:55:17     & $-$03:10:39  & 13.3    &  805&2			  &  3&59\\
     \hline
     \multicolumn{8}{l}{RXJ1730 = \object{1RXS~J173021.5$-$055933},}\\
     \multicolumn{8}{l}{RXJ2133 = \object{1RXS~J213344.1$+$510725}.}\\
   \end{tabular}
\end{table}

\begin{table*}
  \centering
    \caption{Observing log.}
    \label{observing_log}
    \begin{tabular}{lllr@{.}lllllll}
      \hline\hline
      Name      & Start time   & End time     & \multicolumn{2}{l}{Duration} & No. of P$_{\rm{orb}}$ & No. of P$_{\rm{spin}}$ & Filters     & Exposure time$^b$ & Resolution$^c$ & V$^d$\\
                & (HJD$^a$)    & (HJD$^a$)    & \multicolumn{2}{l}{(mins)}   &		      	 &		      &             & (s)	          & (s)            & (mag) \\
      \hline
      DQ Her    & 13\,948.4240 & 13\,948.5349 & 159&7		       & 0.57		 & 67.4                   & {\sl UBVRI} & $1\times10$       & $\sim24$       & 14.5\\
      RXJ2133   & 13\,948.5732 & 13\,948.6542 & 116&6		       & 0.27		 & 12.3		      & {\sl UBVRI} & $4\times10$       & $\sim96$       & 15.3\\
      FO Aqr    & 13\,948.6933 & 13\,948.7091 &  22&8                        & 0.08	 	 & 1.1$^e$                & {\sl UBVRI} & $2\times10$       & $\sim48$       & 13.9\\
      AE Aqr    & 13\,949.5103 & 13\,949.5704 &  86&5                        & 0.15		 & 156.8		      & {\sl UBVRI} & $1\times3$        & $\sim8.5$      & 11.4\\
      V2306 Cyg & 13\,949.5794 & 13\,949.6528 & 105&7                        & 0.40		 & 4.3                    & {\sl UBVRI} & $2\times10$	& $\sim48$       & 14.7\\
      AO Psc    & 13\,949.7047 & 13\,949.7194 &  21&2                        & 0.10 		 & 1.6$^e$                & {\sl UBVRI} & $2\times10$	& $\sim48$       & 13.2\\
      RXJ1730   & 13\,950.4112 & 13\,950.5134 & 147&2                        & 0.16		 & 69.0                   & {\sl UBVRI} & $1\times10$	& $\sim24$       & 16.3\\
      V1223 Sgr & 13\,950.5298 & 13\,950.5871 &  82&5                        & 0.41		 & 6.6                    & {\sl UBVRI} & $2\times10$	& $\sim48$       & 13.7\\
      AE Aqr    & 13\,950.6224 & 13\,950.6366 &  20&4                        & 0.03		 & 37.0		      & {\sl UBVRI} & $1\times1$        & $\sim4.5$      & 11.6\\
      RXJ2133   & 13\,950.6455 & 13\,950.7187 & 105&4		       & 0.24		 & 11.1		      & {\sl UBVRI} & $4\times10$       & $\sim96$       & 15.2\\
      \hline
      \multicolumn{11}{l}{$^a$ $+$2\,440\,000}\\
      \multicolumn{11}{l}{$^b$ The number of orientations of the wave plate, and the time spent at each orientation.}\\
      \multicolumn{11}{l}{$^c$ The resolution is roughly the number of
      orientations of the wave plate multiplied by the exposure time
      multiplied by two (to account for}\\
      \multicolumn{11}{l}{the ordinary and extraordinary measurements) plus some mechanical dead time.}\\
      \multicolumn{11}{l}{$^d$ Measured.}\\
      \multicolumn{11}{l}{$^e$ These data sets are short and therefore the reported uncertainties are probably underestimated.}\\
    \end{tabular}
\end{table*}

The instrumental polarization was small in all bands (see
Table~\ref{calibration_data}). The circular polarization standard star
\object{GRW+70~8247} \citep{west89} was used to check the calibration,
the values reported here are consistent with previous measurements of
the standard. The uncertainties quoted on each
circular polarization measurement are based on photon noise and are
one sigma.

\begin{table}
    \centering
    \caption{Calibration data (taken on the first night).}
    \label{calibration_data}
    \begin{tabular}{lllll}
        \hline\hline
                & \multicolumn{2}{l}{Instrumentational} & \multicolumn{2}{l}{Measured standard}\\
                & \multicolumn{2}{l}{polarization}      & \multicolumn{2}{l}{polarization}\\
        Band    & Value    & uncertainty		      & Value    & uncertainty\\
                & (\%)     & (\%)		      & (\%)     & (\%)\\
        \hline
        {\sl U} & $-0.005$ & 0.030		      & $+0.126$ & $0.103$\\
        {\sl B} & $-0.064$ & 0.024		      & $-3.607$ & $0.112$\\
        {\sl V} & $-0.017$ & 0.028		      & $-3.959$ & $0.166$\\
        {\sl R} & $-0.011$ & 0.024		      & $-4.064$ & $0.156$\\
        {\sl I} & $-0.059$ & 0.029		      & $-2.647$ & $0.226$\\
        \hline
    \end{tabular}
\end{table}

The zero points of the {\sl UBVRI} magnitude scale were determined by
observations of Landolt standards (109\,954, 111\,250, 111\,2093 and
114\,637; \citet{landolt92}) during each night.

\section{Results}
\label{results}

Successful measurements are outlined for seven of the targets
below. The results of RXJ2133 are reported separately in
\cite{katajainen07}.

Given the ambiguity in the previously reported results, we give our
data in multiple formats. The average value over the run will show any
large unmodulated polarization (like that in BG~CMi) and allow
comparison with most of the previous measurements. The peak amplitude
(of the spin folded and phase binned data)
gives an idea of how modulated the system is. The peak to peak value
shows whether or not both magnetic poles can be seen, and gives the
best indication of the presence of modulation.

\subsection{AE Aqr}

\object{AE~Aqr} has the shortest spin period of all the known IPs ($\sim$33~s), and a
relatively long orbital period of 9.88~h. Observed over two nights it
was at an orbital phase of 0.6 and 0.3 from superior conjunction of
the WD with respect to the secondary on the two nights respectively \citep{welsh93}.
We used a spin period of 0.000382833~d, calculated for July/August 2006
from the ephemeris of \citet{dejager94}. The zero spin phase point used here
is arbitrarily set to the midnight epoch at HJD~2\,453\,949.5.

Over the two nights this system showed a marked difference in its
behaviour. During the first night the {\sl U} band exhibited significant flickering
and a general trend of an increase in magnitude (see
Fig.~\ref{ae_aqr_raw_fot}). This was mirrored in
the raw {\sl U} band circular
polarization where the magnitude increased as the run went on. On the
second night the {\sl U} band exhibited
significantly less flickering. This was a trend that was seen in all
bands to some extent (see Fig.~\ref{ae_aqr_raw_fot}). This flickering
is a well known feature of AE~Aqr (see e.g. \citet{beskrovnaya96}).
With this in mind the data from the two nights is presented separately.

\onlfig{1}
{
  \begin{figure}
    \begin{minipage}[t]{.24\textwidth}
      \vspace{0pt}
      \centering
      \includegraphics[height=3.5in]{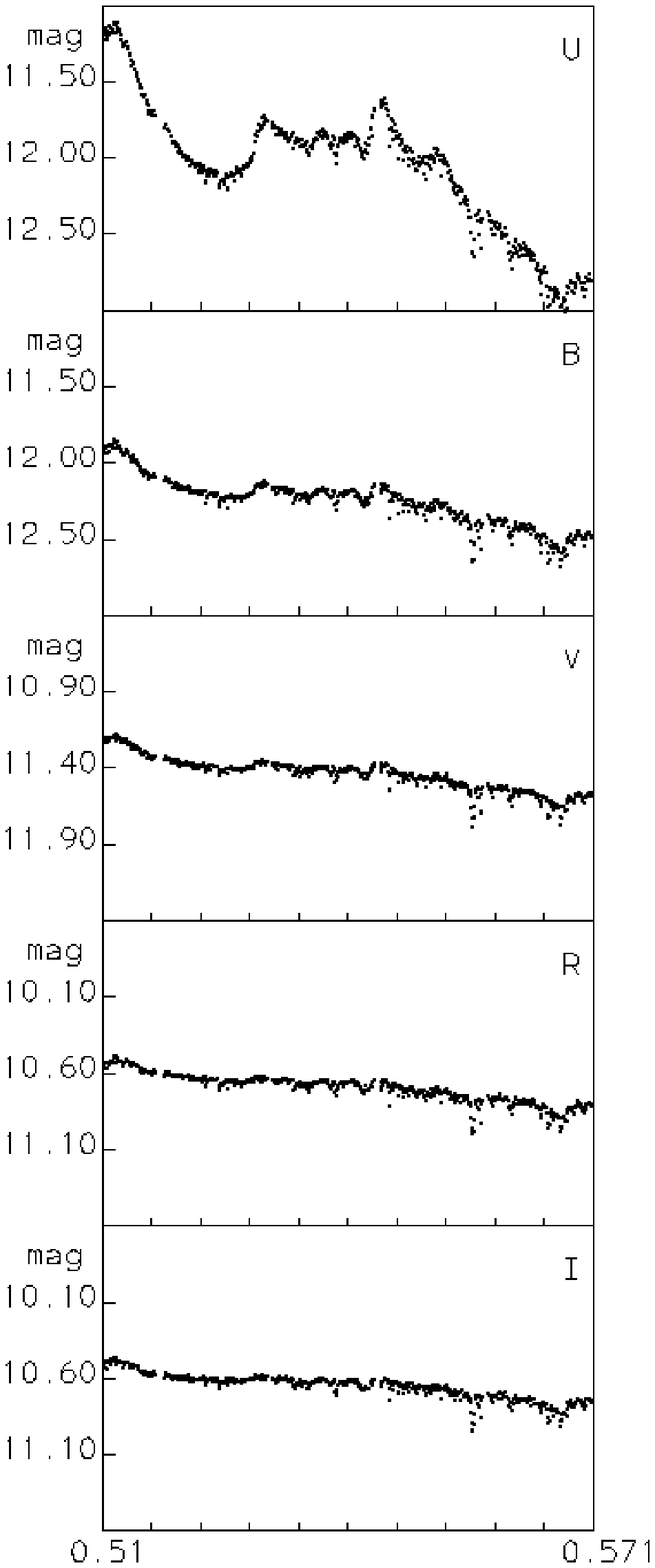}
    \end{minipage}
    \begin{minipage}[t]{.24\textwidth}
      \vspace{0pt}
      \centering
      \includegraphics[height=3.5in]{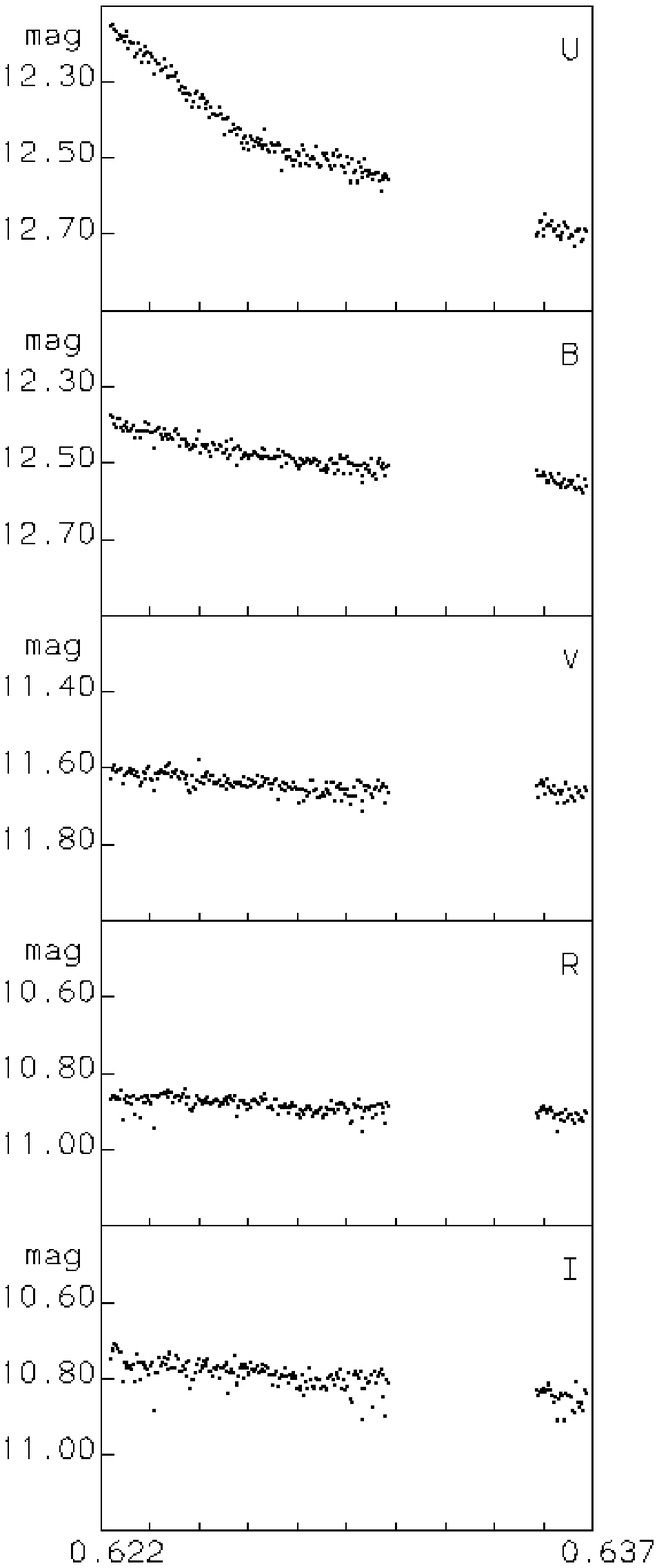}
    \end{minipage}
    \caption{Raw {\sl UBVRI} photometry of AE~Aqr taken on the two
      nights. The abscissa is in left plot is HJD -- 2\,453\,949, in the
      right plot it is HJD -- 2\,453\,950.}
    \label{ae_aqr_raw_fot}
  \end{figure}
}

\begin{figure}
  \begin{minipage}[t]{.24\textwidth}
    \vspace{0pt}
    \centering
    \includegraphics[height=3.5in]{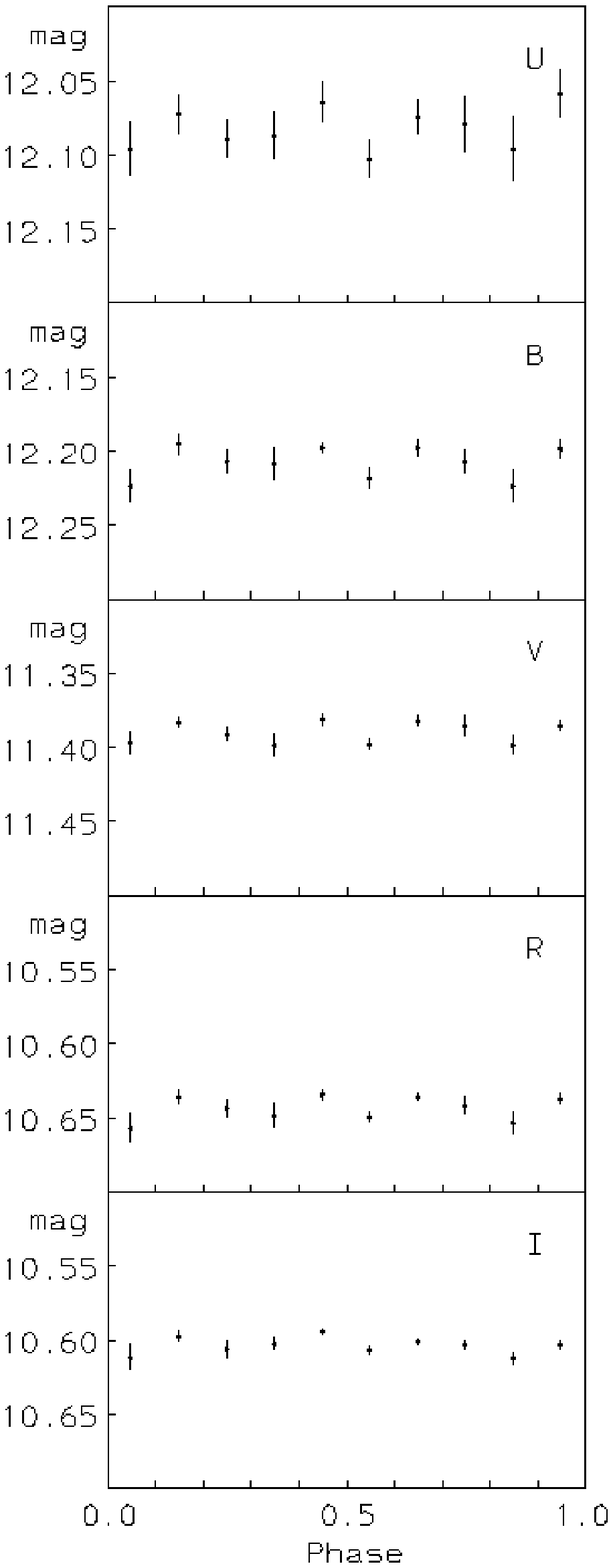}
\end{minipage}
\begin{minipage}[t]{.24\textwidth}
    \vspace{0pt}
    \centering
    \includegraphics[height=3.5in]{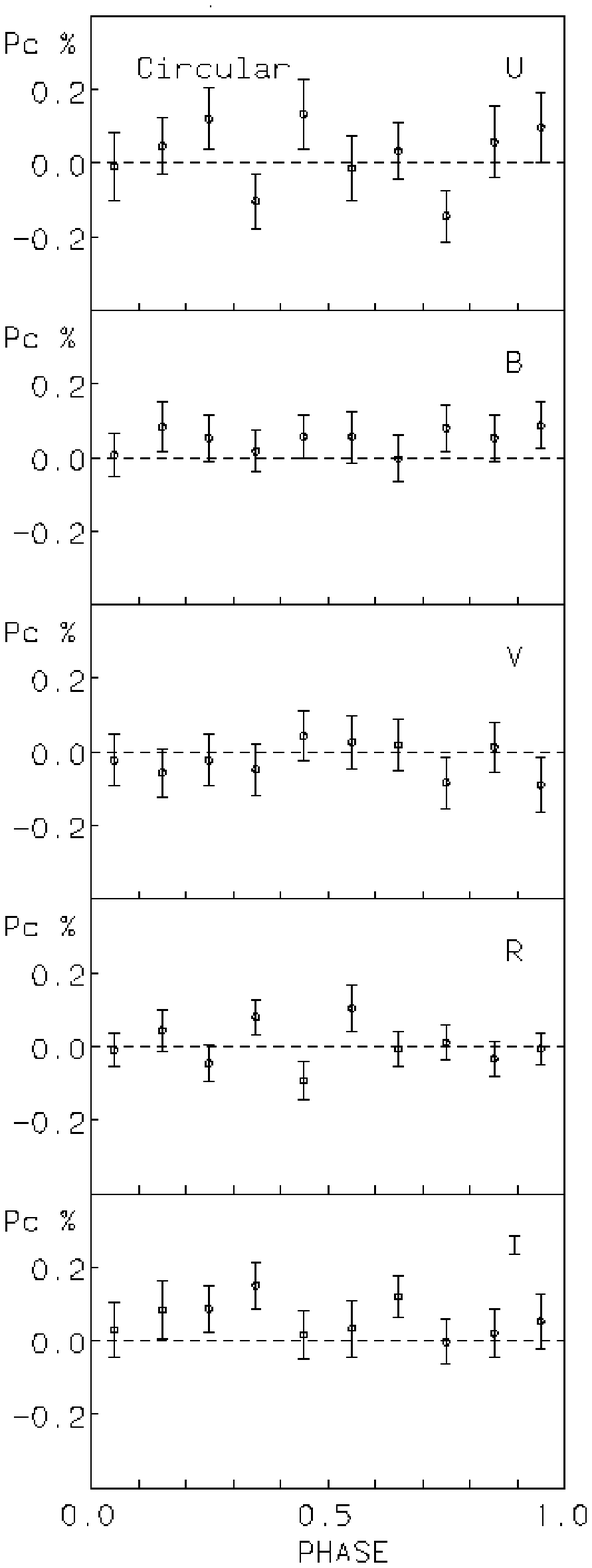}
  \end{minipage}
  \caption{Spin folded and phase binned simultaneous {\sl UBVRI} photometry (left)
and circular polarization (right)
  plots of AE~Aqr, taken on the first night (HJD~2\,453\,949). Zero phase corresponds to
  HJD~2\,453\,949.5. A spin period of 0.000382833 d was used.}
  \label{ae_aqr_fot_cp_night1}
\end{figure}

The short period of the system is such that the temporal resolution
must be as small as possible to search for any periodic variations. In order to satisfy this only one
position of the wave plate was used for each polarization
measurement. On the first night an integration time of
$\sim3$~s was used, this gave an overall temporal resolution of
$\sim8.5$~s. On
the second night an integration time of $\sim1$~s was used, giving a
temporal resolution of $\sim4.5$~s for a full polarization measurement. Even at this short time scale the
measurements will be smoothed to some extent. The data was folded and
binned into 10 bins over the spin cycle (see Figs.~\ref{ae_aqr_fot_cp_night1} and
\ref{ae_aqr_fot_cp_night2}). Both nights show a very small amplitude circular
polarization. However in the raw data, values with an amplitude of over
2\% (with a typical error of 0.6\%) are not uncommon.

The mean values in each band (over each of the nights) are all within three sigma
of zero. The peak amplitude is $0.80\pm0.39\%$ and the maximum peak to
peak value is $1.22\pm0.48\%$ (see Table~\ref{results_summary}). The short spin period means that this data set covers many spin
periods, and therefore the signal to noise is good.

\onlfig{3}{
  \begin{figure}
    \begin{minipage}[t]{.24\textwidth}
      \vspace{0pt}
      \centering
      \includegraphics[height=3.5in]{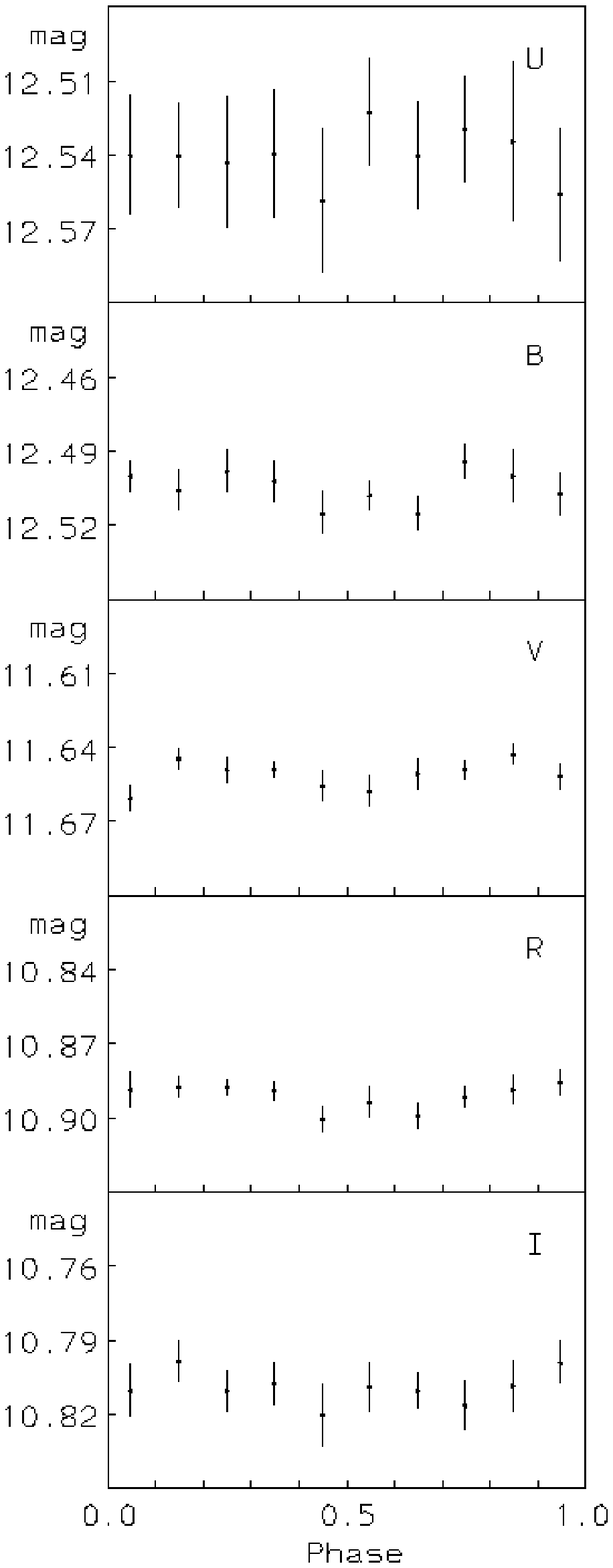}
    \end{minipage}
    \begin{minipage}[t]{.24\textwidth}
      \vspace{0pt}
      \centering
      \includegraphics[height=3.5in]{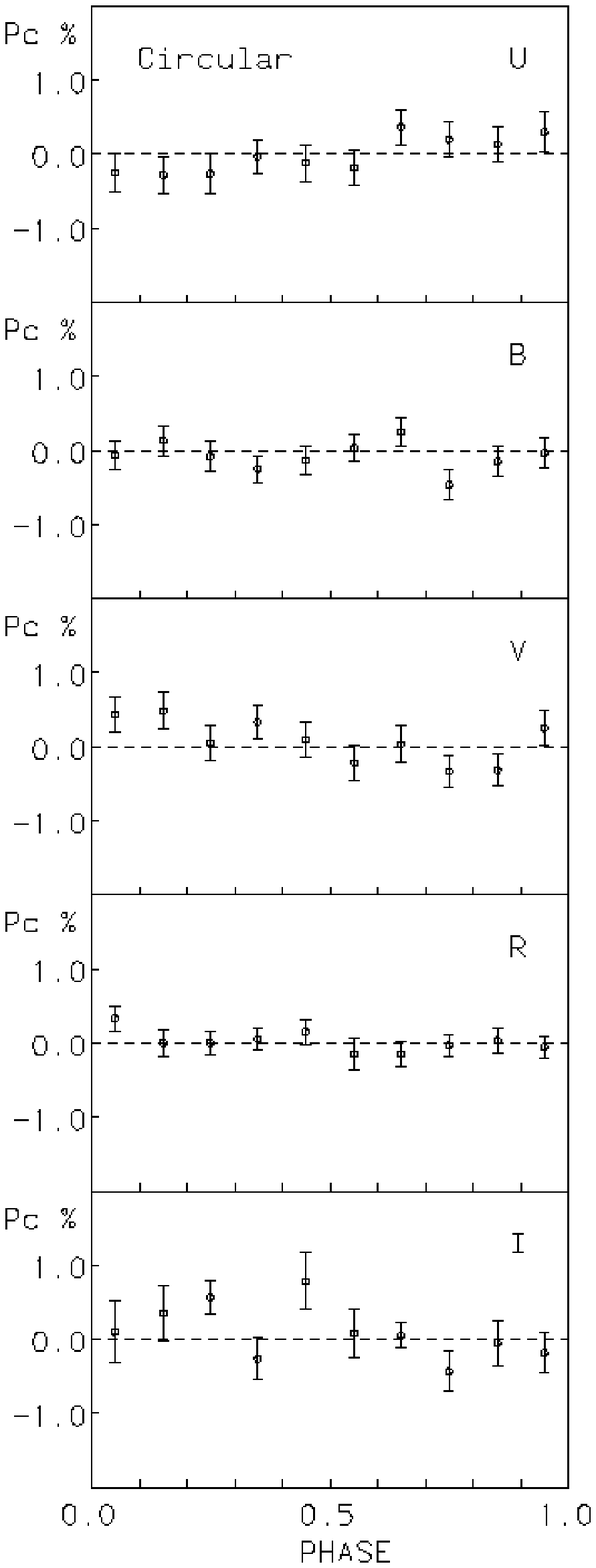}
    \end{minipage}
    \caption{Spin folded and phase binned simultaneous {\sl UBVRI} photometry (left) and circular polarization (right)
      plots of AE~Aqr, taken on the second night (HJD~2\,453\,950). Zero phase corresponds to
      HJD~2\,453\,949.5. A spin period of 0.000382833 d was used.}
    \label{ae_aqr_fot_cp_night2}
  \end{figure}
}

\subsection{AO Psc}
\object{AO~Psc} is a typical IP with a spin and orbital period of 805.2~s and
3.59~h respectively. Using the orbital ephemeris of \citet{kaluzny88}, AO~Psc was at an orbital phase of
0.14 from the maximum optical light. We note that the error in the
calculation of this phase is small, but the ephemeris is old
(20~years) so it may be somewhat out of date. The spin ephemeris has an accumulated uncertainty of greater than one
spin, so we have used a zero point of HJD~2\,453\,949.5. The spin
period from \citet{kaluzny88} of 0.0009319484 d was used.

Each polarization measurement consisted of two positions of the
wave plate at 10~s each. The polarization data was then binned into eight bins
across the spin cycle. The mean values in each band were within three sigma of
being zero. The maximum amplitude in the
binned data was $0.86\pm0.37$\%, with a maximum peak to peak
variation of $1.30\pm0.50$\%, see Fig.~\ref{ao_psc_cp_fot} and Table~\ref{results_summary}. This
peak to peak variation is less than three sigma, so we
cannot claim this as a reliable detection of variable
polarization.

\onlfig{4}
{
  \begin{figure}
    \begin{minipage}[t]{.24\textwidth}
      \vspace{0pt}
      \centering
      \includegraphics[height=3.5in]{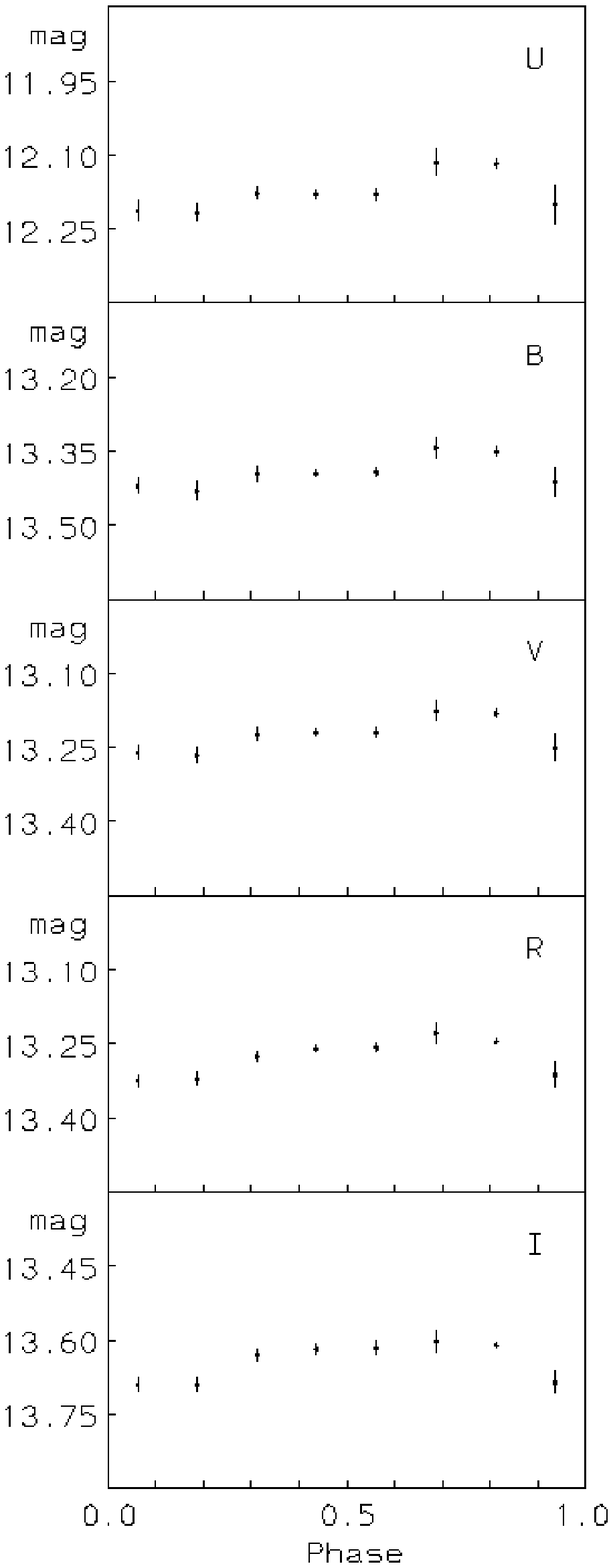}
    \end{minipage}
    \begin{minipage}[t]{.24\textwidth}
      \vspace{0pt}
      \centering
      \includegraphics[height=3.5in]{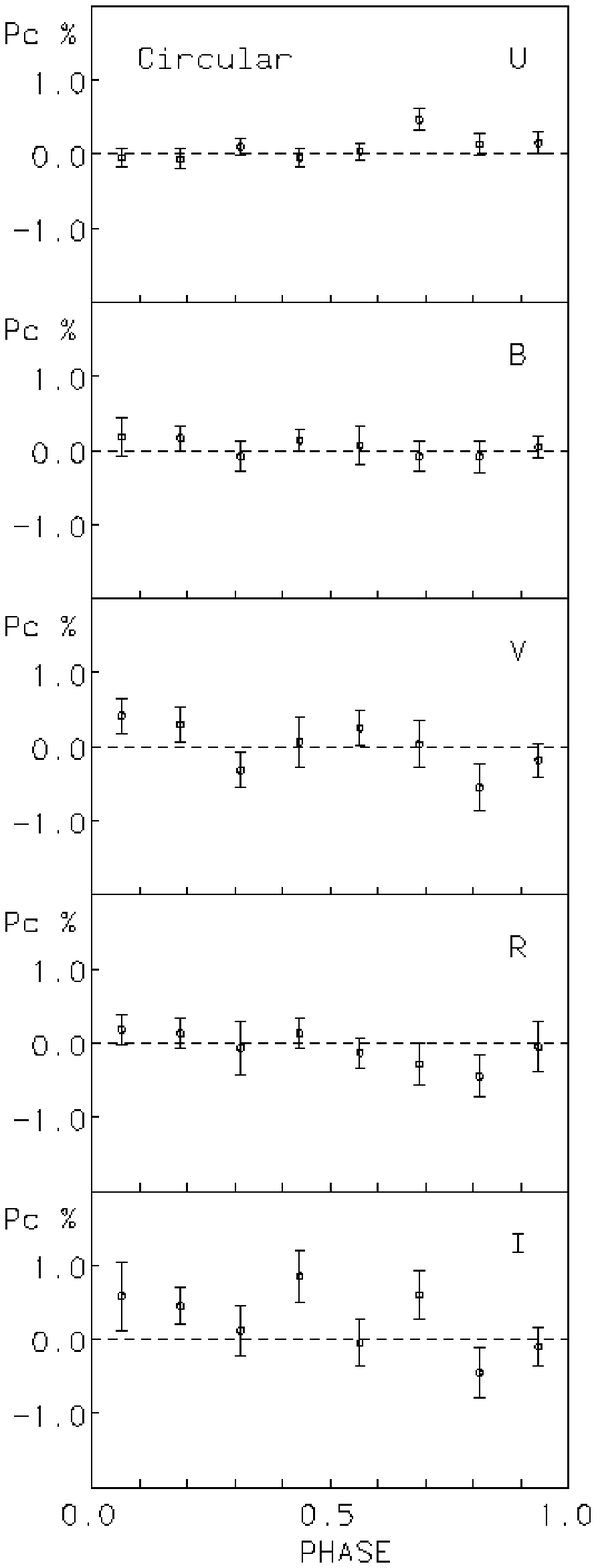}
    \end{minipage}
    \caption{Spin folded and phase binned simultaneous {\sl UBVRI} photometry (left) and circular polarization (right)
      plots of AO~Psc. Circular polarization measurements have an 23~s resolution. Zero phase corresponds to HJD~2\,453\,949.5. A spin period of 0.009319484~d was used.
    }
    \label{ao_psc_cp_fot}
  \end{figure}
}

\subsection{DQ Her}

\object{DQ~Her} has an orbital period of 4.65~h and a spin period of
142~s. Using the ephemerids of \citet{zhang95} \object{DQ~Her} was at
an orbital phase of 0.23 from the optical eclipse. We used an arbitrary zero
point of phase as HJD~2\,453\,948.5. A spin value of 0.00164504~d was
used.

Due to the short period, a single wave plate position was used for an
integration time of 10~s. The data was binned into ten bins over the spin period.
All of the mean polarization values are less than three sigma from zero. The
maximum amplitude seen is $0.64\pm0.31\%$ and the maximum peak to peak
value was $1.00\pm0.39\%$ (see Table~\ref{results_summary}), this is less than a three sigma detection
of variation (see Fig.~\ref{dq_her_cp_fot}).

\onlfig{5}
{
  \begin{figure}
    \begin{minipage}[t]{.24\textwidth}
      \vspace{0pt}
      \centering
      \includegraphics[height=3.5in]{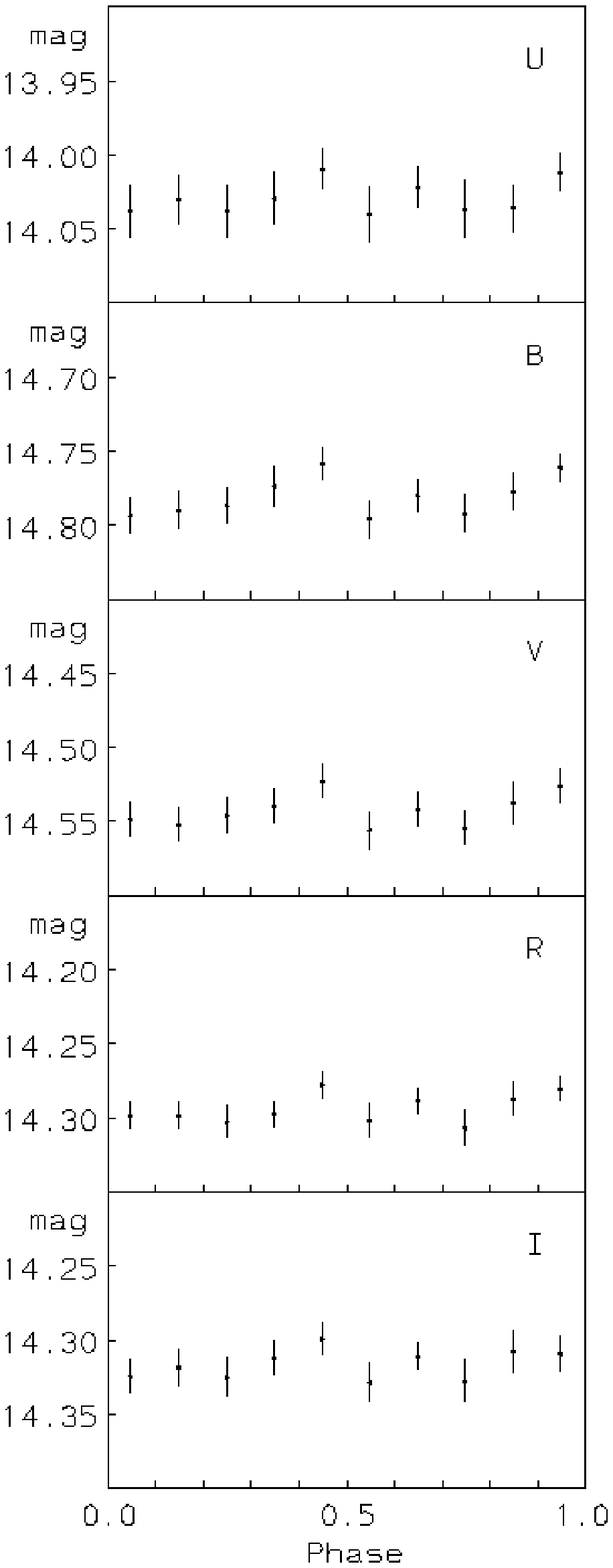}
    \end{minipage}
    \begin{minipage}[t]{.24\textwidth}
      \vspace{0pt}
      \centering
      \includegraphics[height=3.5in]{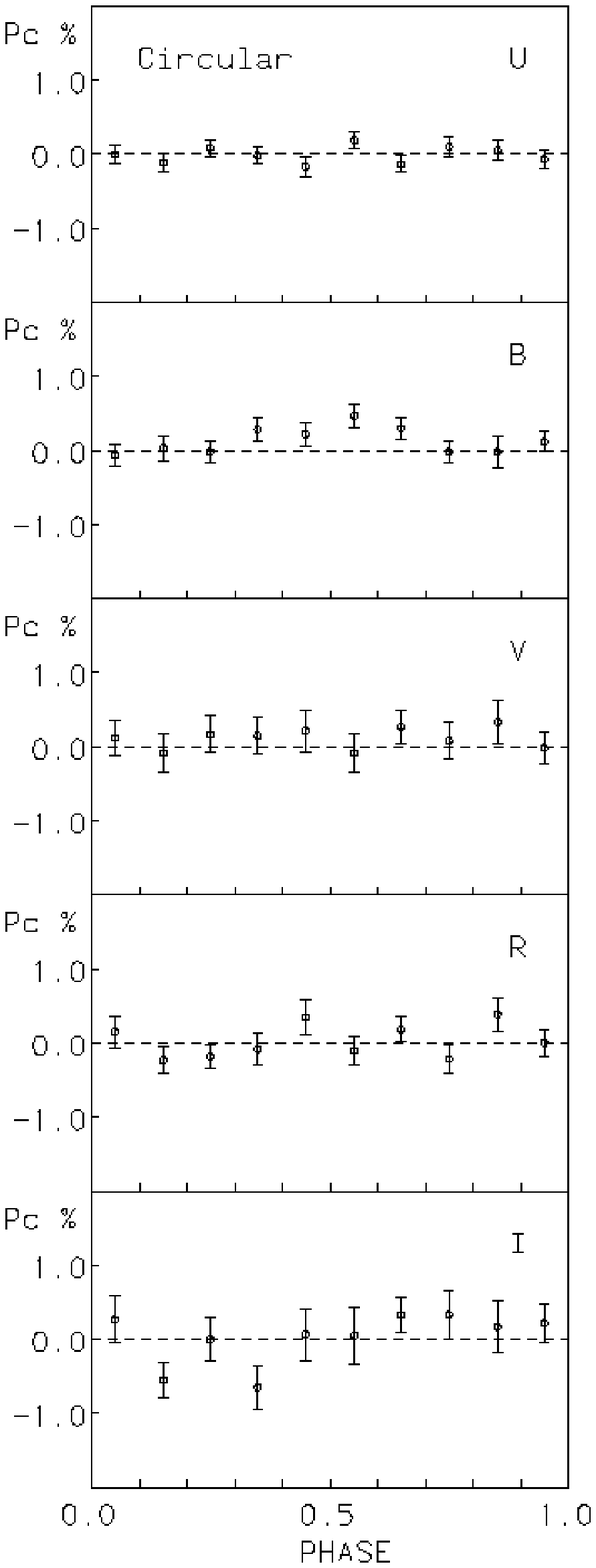}
    \end{minipage}
    \caption{Spin folded and phase binned simultaneous {\sl UBVRI} photometry (left) and circular polarization (right)
      plots of DQ~Her. Zero phase corresponds to HJD~2\,453\,948.5. A spin value of 0.00164504~d was used.
    }
    \label{dq_her_cp_fot}
  \end{figure}
}

\subsection{FO Aqr}
\object{FO~Aqr} has an orbital period of 4.85~h and a spin period of
1254.5~s. Using the ephemeris of \citet{patterson98}, FO~Aqr was at an orbital
phase of 0.98 from the dip in the optical light curve at the start of
this observation. It was also at a spin phase of approximately 0.6 from
pulse maximum, but we note that FO~Aqr is rather erratic and this
value may be some way off now, so zero phase was arbitrarily set to
HJD~2\,453\,948.5. Here a spin value of 0.014519035~d from
\citet{patterson98} was used.

Each polarization measurement was taken with two positions of the
wave plate with an integration time of 10~s in each.
The data was binned into four bins over the spin cycle. The mean circular
polarization was within two sigma of being zero in each band. A peak
amplitude of $1.15\pm0.65\%$ is present in the {\sl I} band (see
Fig.~\ref{fo_aqr_cp_fot}). The peak to peak values had a maximum of
$1.43\pm0.80\%$ (see Table~\ref{results_summary}).

\onlfig{6}
{
  \begin{figure}
    \begin{minipage}[t]{.24\textwidth}
      \vspace{0pt}
      \centering
      \includegraphics[height=3.5in]{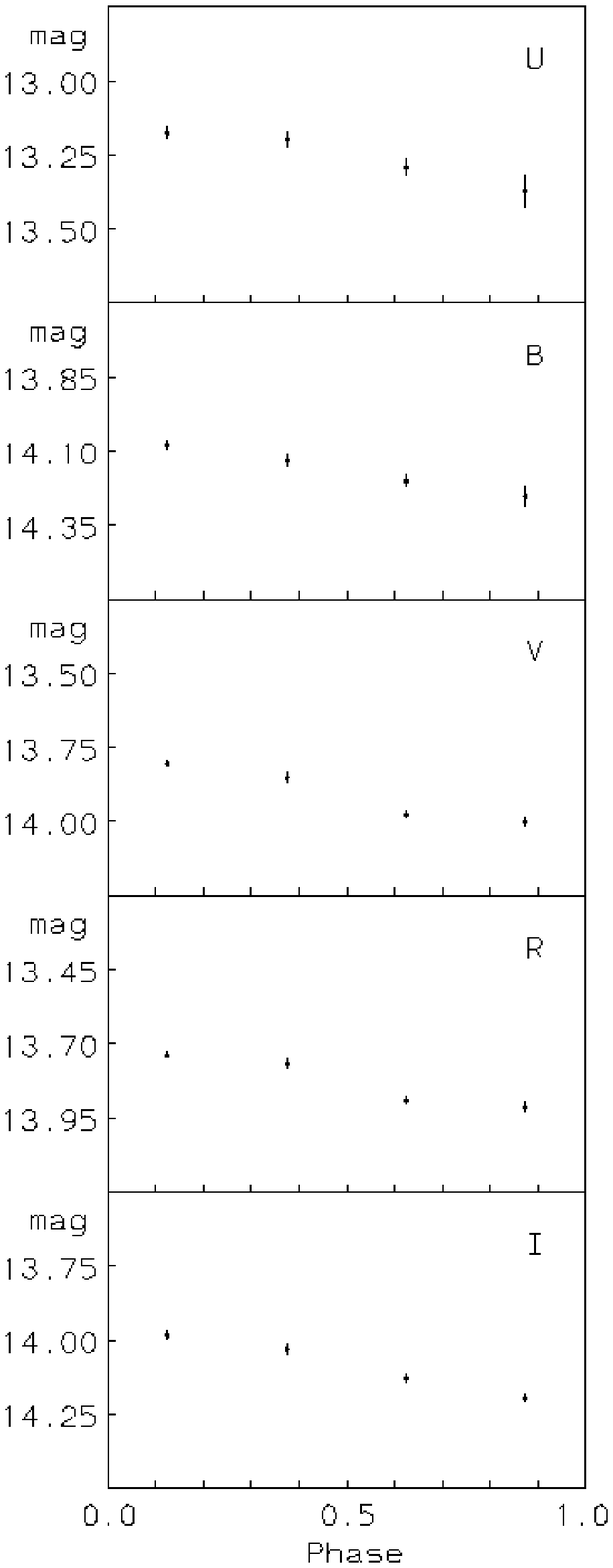}
    \end{minipage}
    \begin{minipage}[t]{.24\textwidth}
      \vspace{0pt}
      \centering
      \includegraphics[height=3.5in]{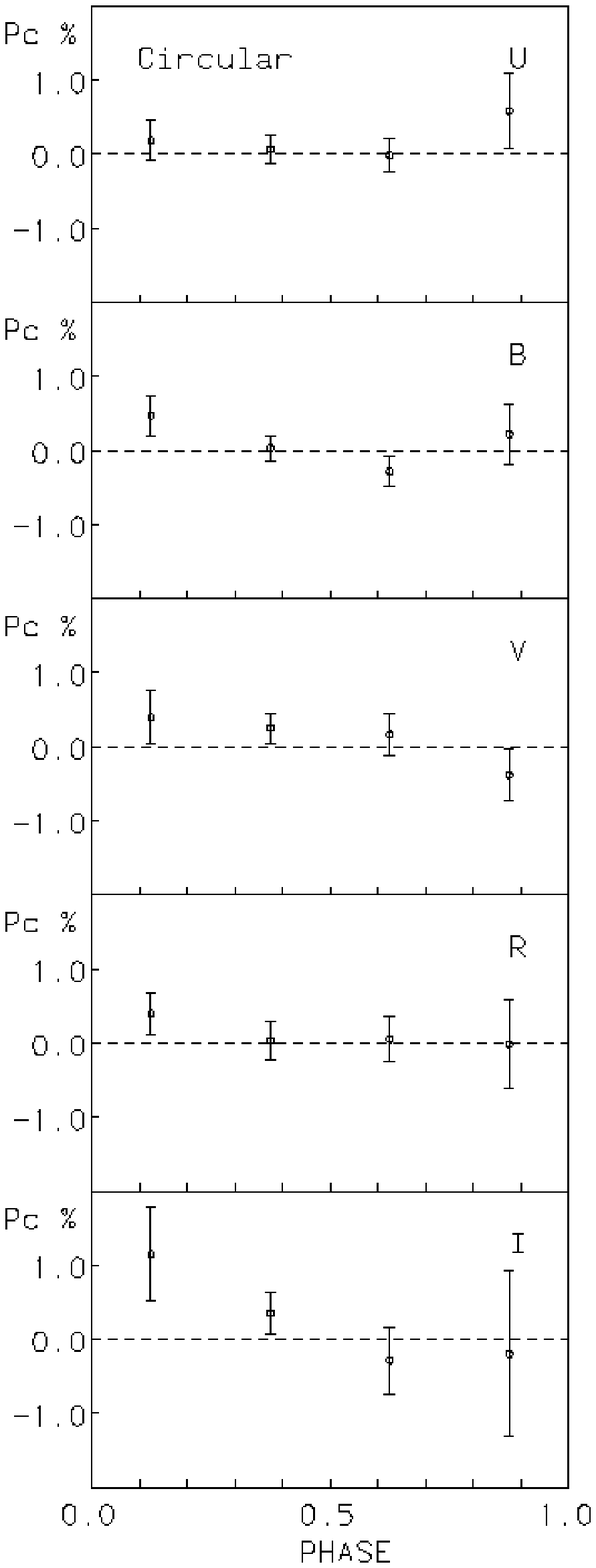}
    \end{minipage}
    \caption{Spin folded and phase binned simultaneous {\sl UBVRI} photometry (left) and circular polarization (right)
      plots of FO~Aqr. Zero phase corresponds to HJD~2\,453\,948.5. A spin period of
      0.0014519035~d was used.
    }
    \label{fo_aqr_cp_fot}
  \end{figure}
}

\subsection{1RXS~J173021.5--055933}
\object{1RXS~J173021.5--055933} (RXJ1730) is a relatively newly classified
IP. It has a reported orbital period of 15.42~hr and spin period of
128.0~s \citep{gansicke05}. This short period (and therefore large
number of spin cycles) has rendered the spin ephemeris of \citet{demartino08} out
of date, and there is no published orbital ephemeris. Zero spin phase
was set to HJD~2\,453\,950.5. A spin period of 0.001481481~d
\citep{gansicke05} was used.

\begin{figure}
  \begin{minipage}[t]{.24\textwidth}
    \vspace{0pt}
    \centering
    \includegraphics[height=3.5in]{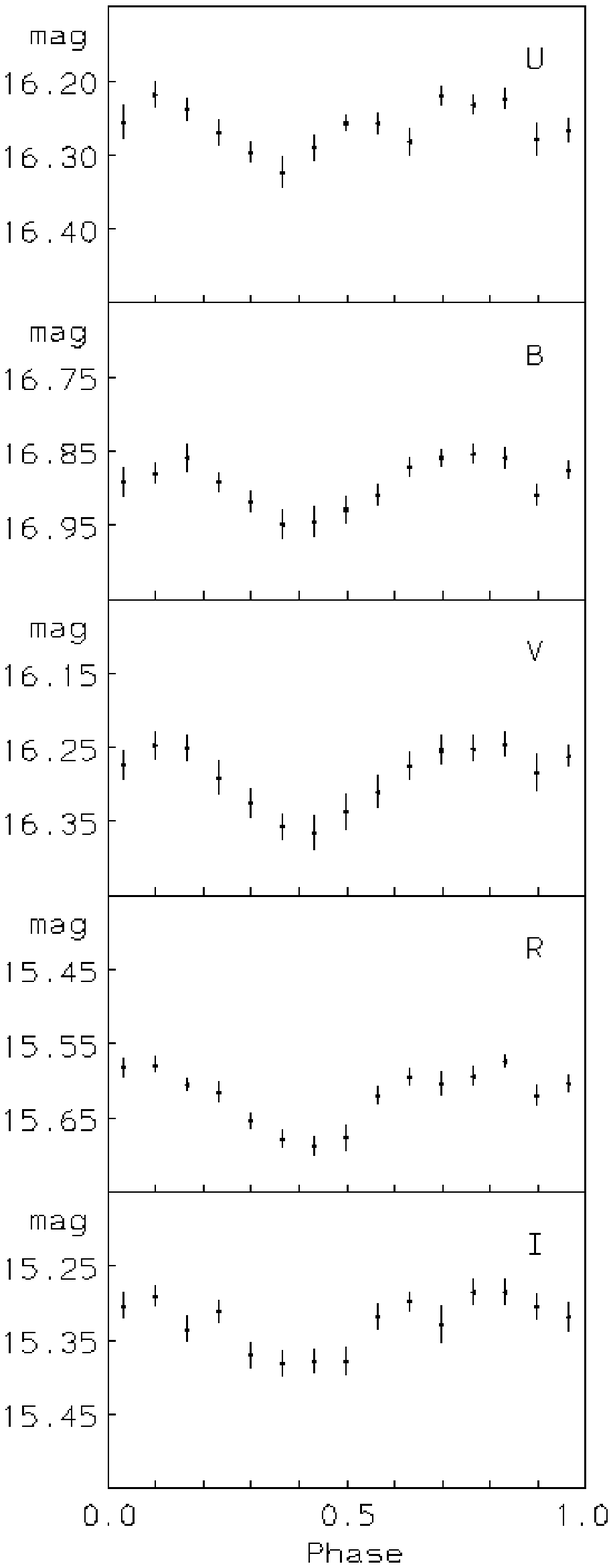}
\end{minipage}
\begin{minipage}[t]{.24\textwidth}
    \vspace{0pt}
    \centering
    \includegraphics[height=3.5in]{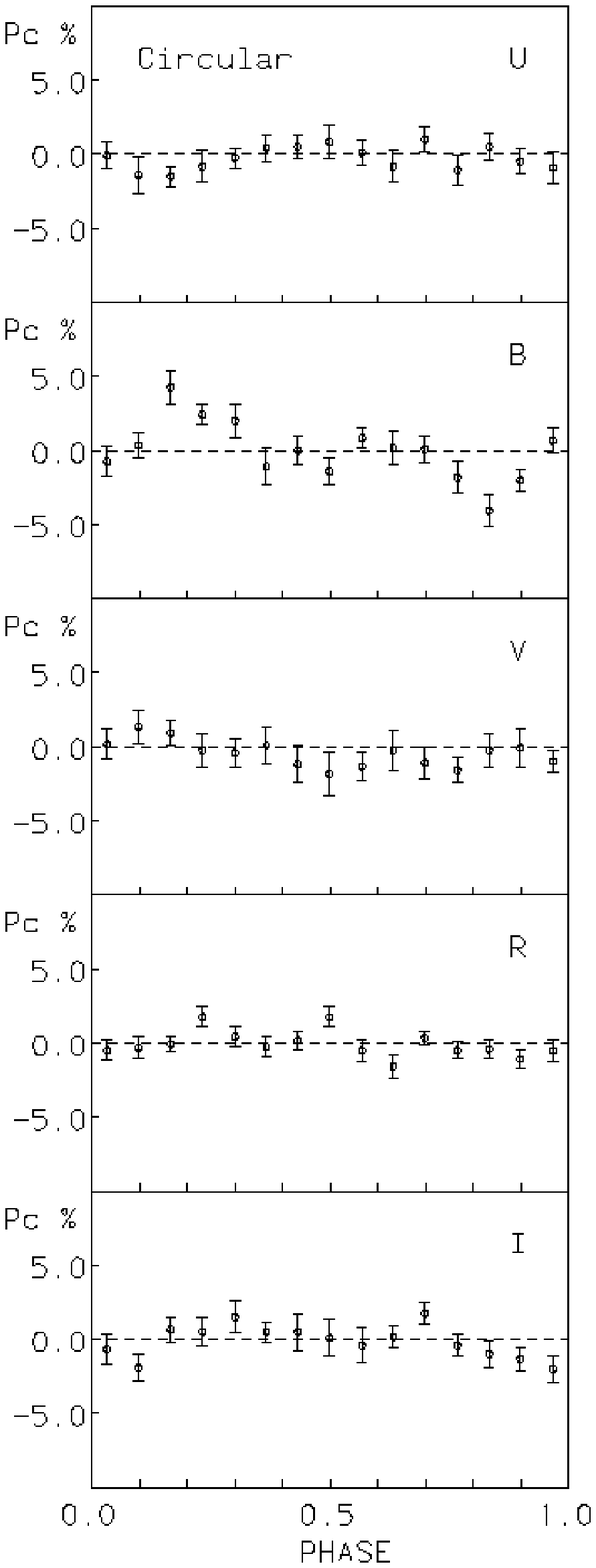}
  \end{minipage}
  \caption{Spin folded and phase binned simultaneous {\sl UBVRI} photometry (left) and circular polarization (right)
  plots of RXJ1730. Zero phase corresponds to HJD~2\,453\,950.5. A spin period of 0.001481481~d was used.}
  \label{rxj1730_cp_fot}
\end{figure}

\begin{figure}
  \resizebox{\hsize}{!}{\includegraphics{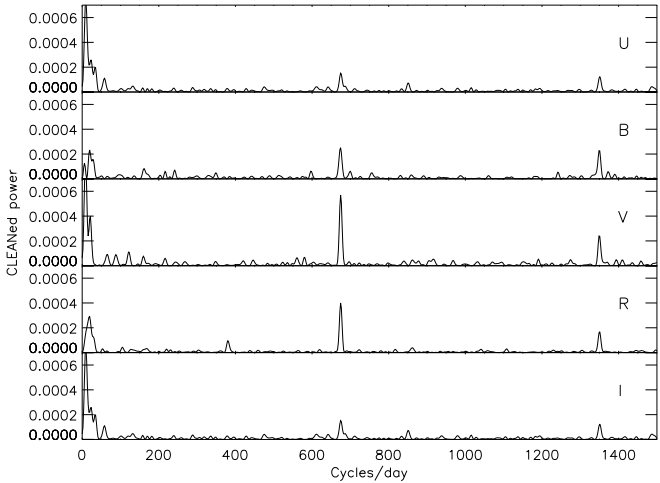}}
  \caption{{\sl UBVRI} {\sc clean}ed photometric periodograms of RXJ1730.}
  \label{rx1730_ubvri_periodograms}
\end{figure}

Here the first simultaneous {\sl UBVRI} photometry of this object is presented (see
Fig.~\ref{rxj1730_cp_fot}). The photometry exhibits a double peak
profile with equal maxima and unequal minima in each band. The
photometric {\sc clean}ed \citep{lehto97} periodograms of each
individual band are shown in Fig.~\ref{rx1730_ubvri_periodograms}. The
spin period is seen at a value of 128.1$\pm$0.7~s and the first
harmonic at 64.0$\pm$0.2~s (uncertainties based on a one sigma Gaussian fit
to the periodogram). The spin peak is seen strongest in the {\sl V} band.

The wave plate was positioned in one orientation for each circular
polarization measurement for 10~s. The data were binned into
15 bins over the spin cycle. In each band
the mean circular polarization was within two sigma of being zero. The
biggest amplitude modulation was $4.26\pm1.09\%$ and the greatest
peak to peak value was $8.26\pm1.56\%$ (see
Table~\ref{results_summary}). The spin period was recovered from a
period search of the {\sl B} band circular polarization also. The
short spin period means that many spin cycles (69) were completed,
this leads to a high confidence in this data set.

\subsection{V1223 Sgr}
\object{V1223~Sgr} has a spin and orbital period of 745.6~s and 3.37~h
respectively. Using the orbital ephemeris of \citet{jablonski87} V1223~Sgr is at an orbital
phase of 0.83 from the maximum light, again this phase is valid with
respect to the ephemeris, but the ephemeris is over 20~years old. The spin ephemeris has accumulated too much uncertainty to
be useful here, so zero phase was arbitrarily set to HJD~2\,453\,950.5. A spin value
of 0.00863~d from \citet{osborne85} was used.

The polarization measurements consisted of two positions of the
wave plate, each of 10~s. The data was binned into ten bins across the spin cycle.
The mean values of the circular polarization are all within three
sigma of being zero. The maximum amplitude variation was $1.30\pm1.12\%$ in
the {\sl U} band and the maximum peak to peak value was
$2.16\pm1.22\%$ (see Fig.~\ref{v1223_cp_fot}).
The peak to peak values are all within three sigma of being zero (see
Table~\ref{results_summary}).

\onlfig{9}
{
  \begin{figure}
    \begin{minipage}[t]{.24\textwidth}
      \vspace{0pt}
      \centering
      \includegraphics[height=3.5in]{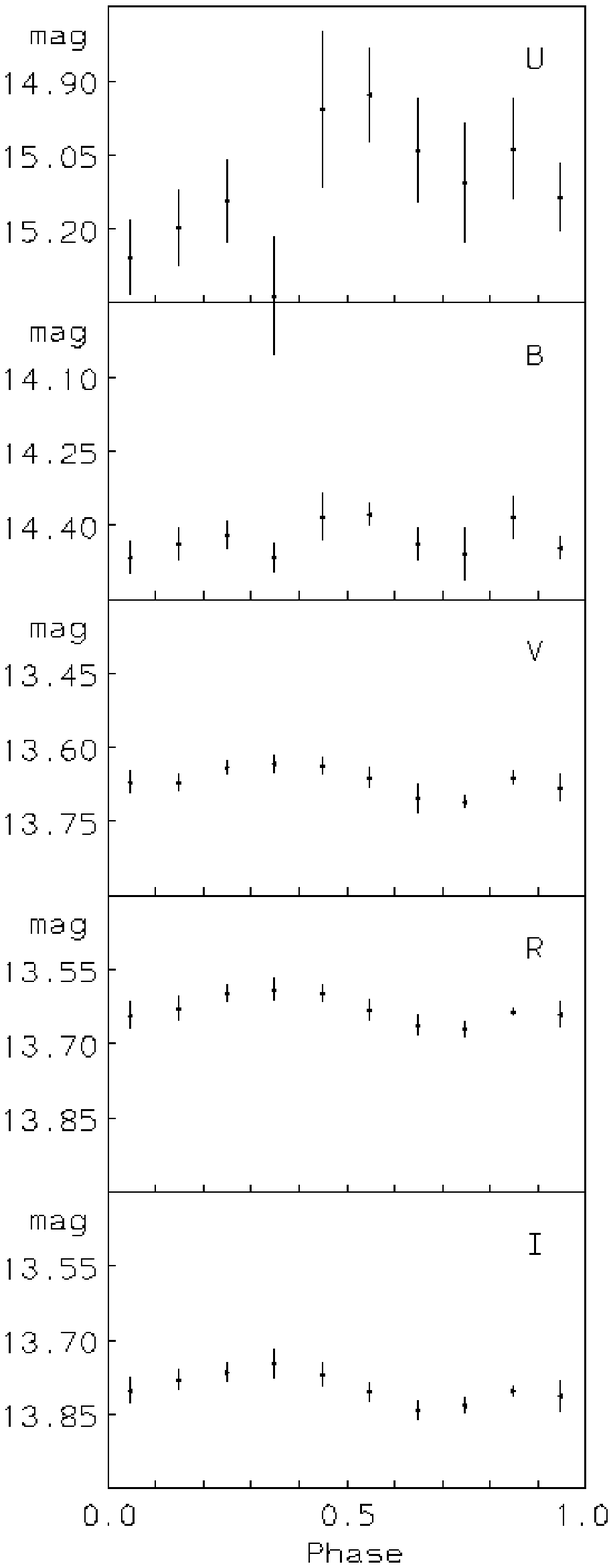}
    \end{minipage}
    \begin{minipage}[t]{.24\textwidth}
      \vspace{0pt}
      \centering
      \includegraphics[height=3.5in]{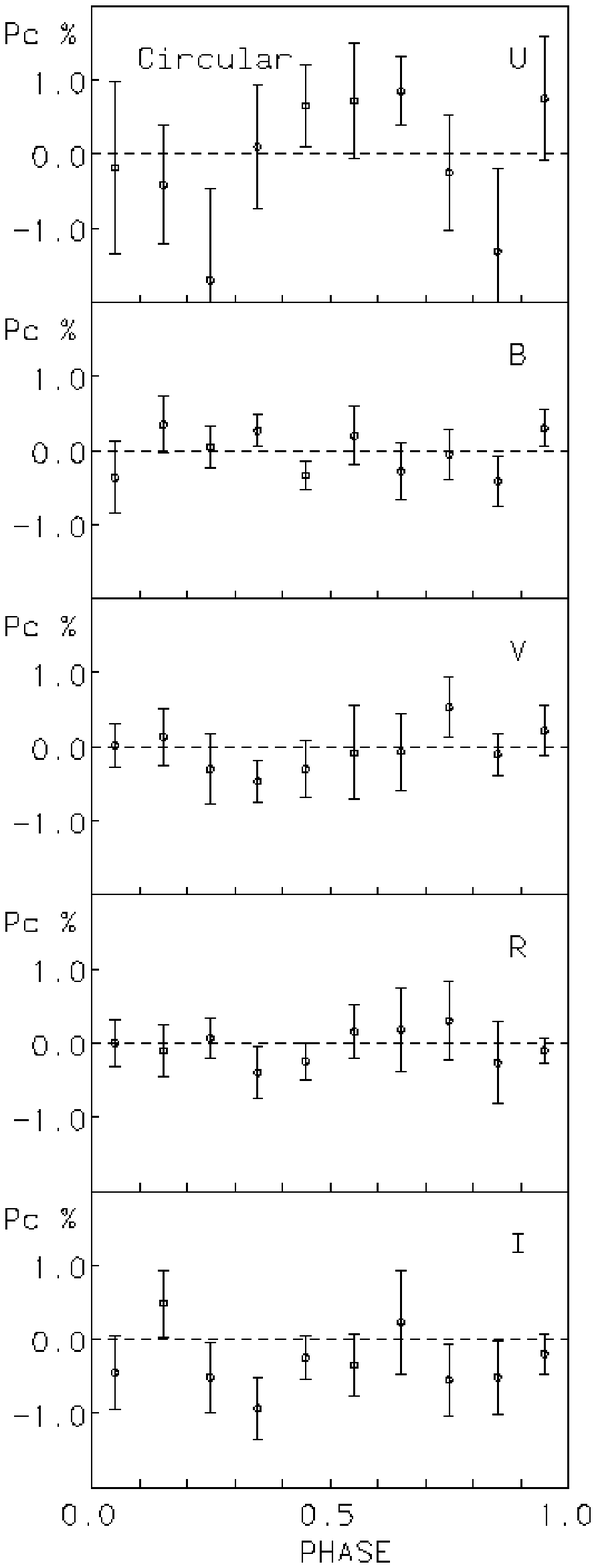}
    \end{minipage}
    \caption{Spin folded and phase binned simultaneous {\sl UBVRI} photometry (left) and circular polarization (right)
      plots of V1223~Sgr. Zero phase corresponds to HJD~2\,453\,950.5. A spin period of 0.00863~d
      was used.
    }
    \label{v1223_cp_fot}
  \end{figure}
}

\subsection{V2306 Cyg}
\object{V2306~Cyg} has an orbital period of 4.35~h and a spin period of
1466.7~s \citep{norton02,zharikov02}. The spin ephemeris of \citet{norton02} was used to phase the spin variations here.

Each polarization measurement consisted of two positions of the
wave plate, each position being 10~s. The data were binned into 15
bins over the spin cycle. The maximum
amplitude circular polarization was $1.06\pm0.41\%$ in the {\sl I} band. The
mean circular polarization in each band is consistent with zero (see
Table~\ref{results_summary}). The maximum peak to peak value
($1.95\pm0.66\%$ in the {\sl I} band) indicates that
variation is present (see Fig.~\ref{v2306_cp_fot} and
Table~\ref{results_summary}). 

\onlfig{10}
{
  \begin{figure}
    \begin{minipage}[t]{.24\textwidth}
      \vspace{0pt}
      \centering
      \includegraphics[height=3.5in]{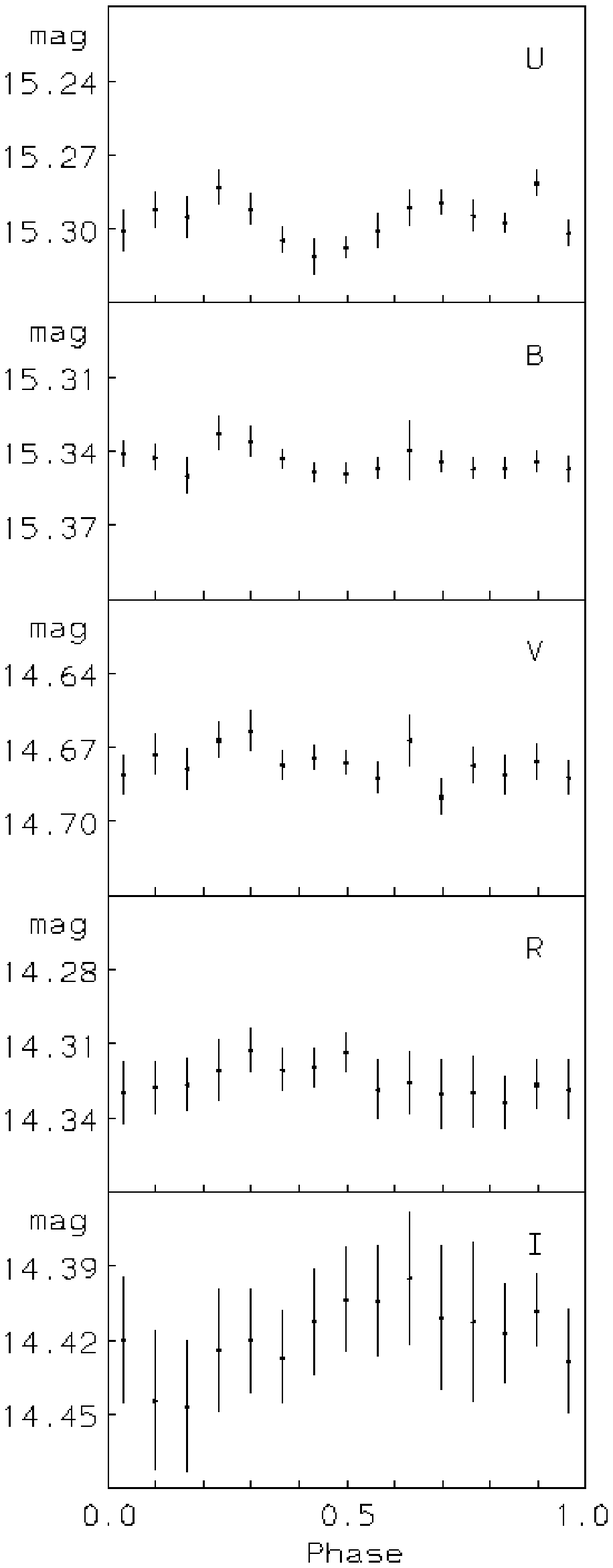}
    \end{minipage}
    \begin{minipage}[t]{.24\textwidth}
      \vspace{0pt}
      \centering
      \includegraphics[height=3.5in]{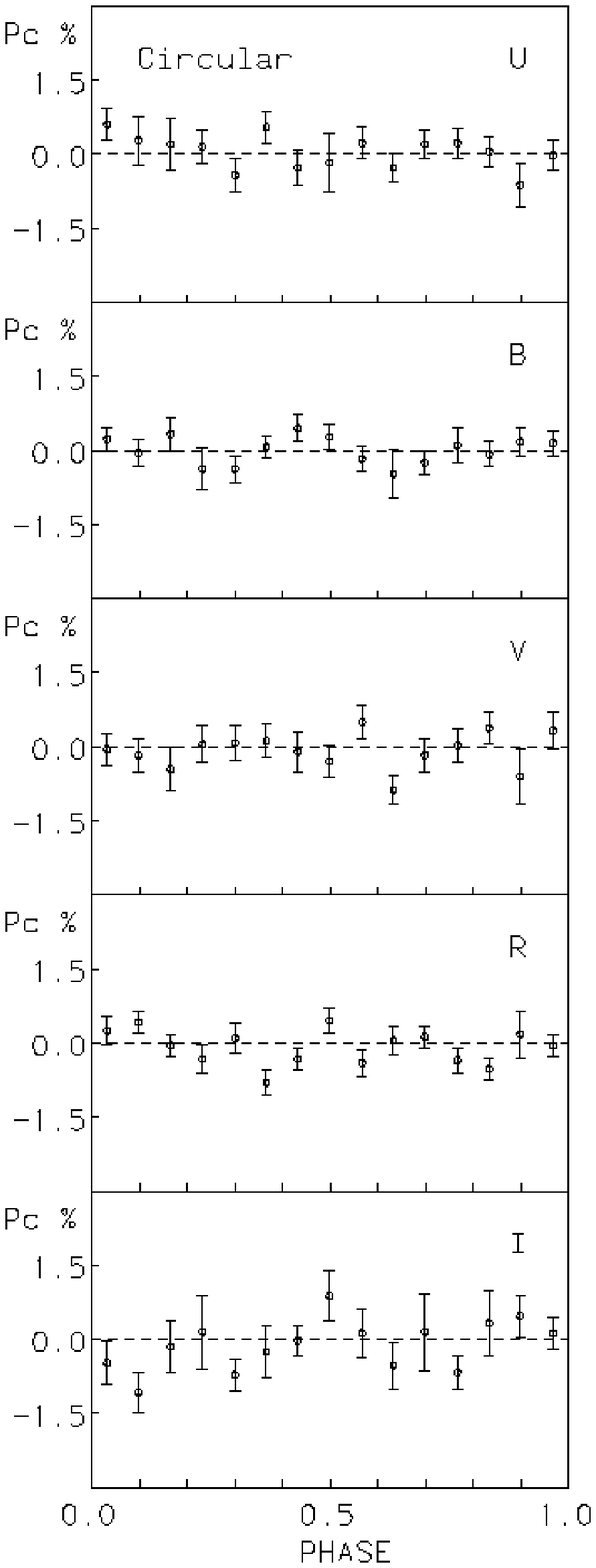}
    \end{minipage}
    \caption{Spin folded and phase binned simultaneous {\sl UBVRI} photometry (left) and circular polarization (right)
      plots of V2306~Cyg. The spin ephemeris of \citet{norton02} was used to phase the spin variations.
    }
    \label{v2306_cp_fot}
  \end{figure}
}

\subsection{Summary of results}

Table~\ref{results_summary} summarises the results obtained, this
should now be used in conjunction with the results in
Table~\ref{already_known_CP_IPs}. As noted earlier, our results
on RXJ2133 have been reported separately in Katajainen et al (2007).

\begin{table*}
  \centering
    \caption{Summary of results. Min and max correspond to the phase binned data.}
    \label{results_summary}
    \begin{tabular}{llllll}
      \hline\hline
      Target       & Filter  & Mean           &  Min            & Max            & Peak-Peak\\
                   &         & (\%)           & (\%)            & (\%)           & (\%)\\
      \hline
      AE Aqr$^{a}$ & {\sl U} & $+0.02\pm0.02$ & $-0.14\pm0.07$ & $+0.13\pm0.10$ & $0.27\pm0.12$\\%from aen1cir_10np_one.pr
                   & {\sl B} & $+0.05\pm0.02$ & $-0.00\pm0.06$ & $+0.09\pm0.06$ & $0.09\pm0.08$\\
                   & {\sl V} & $-0.02\pm0.02$ & $-0.09\pm0.08$ & $+0.04\pm0.07$ & $0.13\pm0.11$\\
                   & {\sl R} & $+0.01\pm0.01$ & $-0.09\pm0.05$ & $+0.11\pm0.06$ & $0.20\pm0.08$\\
                   & {\sl I} & $+0.06\pm0.02$ & $-0.00\pm0.06$ & $+0.15\pm0.07$ & $0.15\pm0.09$\\
      AE Aqr$^{b}$ & {\sl U} & $+0.00\pm0.08$ & $-0.27\pm0.26$ & $+0.38\pm0.23$ & $0.65\pm0.35$\\%from aen2cir_10np_one.pr
                   & {\sl B} & $-0.07\pm0.07$ & $-0.46\pm0.21$ & $+0.26\pm0.21$ & $0.72\pm0.30$\\
                   & {\sl V} & $+0.08\pm0.07$ & $-0.33\pm0.23$ & $+0.50\pm0.26$ & $0.83\pm0.35$\\
                   & {\sl R} & $+0.02\pm0.04$ & $-0.14\pm0.18$ & $+0.34\pm0.18$ & $0.48\pm0.25$\\
                   & {\sl I} & $+0.10\pm0.07$ & $-0.42\pm0.28$ & $+0.80\pm0.39$ & $1.22\pm0.48$\\
      AO Psc       & {\sl U} & $+0.07\pm0.05$ & $-0.06\pm0.14$ & $+0.48\pm0.16$ & $0.54\pm0.21$\\%from 8 bin data
                   & {\sl B} & $+0.07\pm0.06$ & $-0.08\pm0.20$ & $+0.20\pm0.28$ & $0.28\pm0.34$\\
                   & {\sl V} & $+0.05\pm0.09$ & $-0.55\pm0.33$ & $+0.43\pm0.25$ & $0.98\pm0.41$\\
                   & {\sl R} & $-0.02\pm0.08$ & $-0.44\pm0.29$ & $+0.19\pm0.21$ & $0.63\pm0.36$\\
                   & {\sl I} & $+0.29\pm0.10$ & $-0.44\pm0.35$ & $+0.86\pm0.37$ & $1.30\pm0.50$\\
      DQ Her       & {\sl U} & $+0.00\pm0.04$ & $-0.16\pm0.14$ & $+0.19\pm0.13$ & $0.35\pm0.19$\\%from 10 bin data
                   & {\sl B} & $+0.14\pm0.05$ & $-0.05\pm0.17$ & $+0.47\pm0.16$ & $0.53\pm0.23$\\
                   & {\sl V} & $+0.12\pm0.07$ & $-0.08\pm0.25$ & $+0.35\pm0.29$ & $0.43\pm0.39$\\
                   & {\sl R} & $+0.03\pm0.06$ & $-0.22\pm0.20$ & $+0.40\pm0.23$ & $0.61\pm0.31$\\
                   & {\sl I} & $+0.04\pm0.07$ & $-0.64\pm0.31$ & $+0.35\pm0.25$ & $1.00\pm0.39$\\
      FO Aqr       & {\sl U} & $+0.16\pm0.10$ & $-0.00\pm0.24$ & $+0.56\pm0.53$ & $0.60\pm0.58$\\%from 4 bin data
                   & {\sl B} & $+0.07\pm0.10$ & $-0.27\pm0.20$ & $+0.47\pm0.28$ & $0.74\pm0.34$\\
                   & {\sl V} & $+0.17\pm0.15$ & $-0.38\pm0.36$ & $+0.40\pm0.36$ & $0.78\pm0.51$\\
                   & {\sl R} & $+0.11\pm0.12$ & $-0.01\pm0.60$ & $+0.42\pm0.30$ & $0.43\pm0.67$\\
                   & {\sl I} & $+0.27\pm0.20$ & $-0.28\pm0.47$ & $+1.15\pm0.65$ & $1.43\pm0.80$\\
      RXJ1730      & {\sl U} & $-0.23\pm0.18$ & $-1.48\pm0.72$ & $+1.02\pm0.84$ & $2.50\pm1.11$\\%from 15 bin data
                   & {\sl B} & $+0.00\pm0.21$ & $-4.00\pm1.11$ & $+4.26\pm1.09$ & $8.26\pm1.56$\\
                   & {\sl V} & $-0.39\pm0.21$ & $-1.79\pm1.45$ & $+1.38\pm1.19$ & $3.17\pm1.88$\\
                   & {\sl R} & $-0.07\pm0.13$ & $-1.55\pm0.79$ & $+1.80\pm0.67$ & $3.35\pm1.07$\\
                   & {\sl I} & $-0.12\pm0.19$ & $-1.95\pm0.93$ & $+1.82\pm0.75$ & $3.77\pm1.19$\\
      V1223 Sgr    & {\sl U} & $-0.01\pm0.20$ & $-1.30\pm1.12$ & $+0.86\pm0.48$ & $2.16\pm1.22$\\%from 10 bin data
                   & {\sl B} & $-0.00\pm0.08$ & $-0.40\pm0.34$ & $+0.36\pm0.38$ & $0.77\pm0.51$\\
                   & {\sl V} & $-0.04\pm0.09$ & $-0.45\pm0.25$ & $+0.55\pm0.41$ & $1.00\pm0.48$\\
                   & {\sl R} & $-0.05\pm0.09$ & $-0.40\pm0.36$ & $+0.32\pm0.55$ & $0.71\pm0.65$\\
                   & {\sl I} & $-0.27\pm0.11$ & $-0.92\pm0.43$ & $+0.50\pm0.46$ & $1.42\pm0.63$\\
      V2306 Cyg    & {\sl U} & $+0.06\pm0.08$ & $-0.61\pm0.45$ & $+0.61\pm0.33$ & $1.23\pm0.56$\\%from 15 bin data
                   & {\sl B} & $-0.00\pm0.07$ & $-0.47\pm0.51$ & $+0.46\pm0.29$ & $0.92\pm0.59$\\
                   & {\sl V} & $-0.07\pm0.08$ & $-0.86\pm0.29$ & $+0.51\pm0.37$ & $1.36\pm0.48$\\
                   & {\sl R} & $-0.08\pm0.06$ & $-0.78\pm0.26$ & $+0.45\pm0.26$ & $1.23\pm0.37$\\
                   & {\sl I} & $-0.11\pm0.10$ & $-1.06\pm0.41$ & $+0.89\pm0.52$ & $1.95\pm0.66$\\
      \hline
      \multicolumn{6}{l}{$^a$ First night; $^b$ Second night.}\\
      \end{tabular}
\end{table*}

\section{Discussion}
\label{discussion}

\subsection{AE Aqr}

This is the first simultaneous {\sl UBVRI} polarimetry measurement of
AE~Aqr. In most previous measurements a broad band filter and/or a much too
long integration time has been used (see
Table~\ref{already_known_CP_IPs}). This will have had the effect of
smearing any polarization out to almost zero. In the cases where a short
integration time has been used only a mean value has been reported,
except for \citet{cropper86} where a maximum semi-amplitude of $\sim
0.1\%$ was given.

The data reported here broadly agrees with the previous mean
measurements of close to zero. The small peak to peak values are also
in agreement with this. We do note that there is a hint of
circular polarization in the raw data (over 2\% in places - with a
typical error of 0.6\%).

Given the generally accepted view that AE~Aqr is a propeller system it
seems intuitive to assume that it would have a large magnetic field to
power this regime. \citet{norton08} have shown in their theoretical
modelling that propellers can exist at low magnetic field strengths
when they are spinning sufficiently fast. So a large magnetic field in
AE~Aqr is not necessarily required.

\subsection{AO Psc}
This is the first simultaneous {\sl UBVRI} polarimetric observation of
this target. All previous measurements have had a long integration
time ($\gtrsim 240$~s) when compared to the spin period (805.2~s) (see
Table~\ref{already_known_CP_IPs}), so any variations shorter than this
will have been smeared out effectively.

The mean measured circular polarization values are consistent with
previous measurements (all of which were within one sigma of zero) (see
Table~\ref{already_known_CP_IPs}).

The peak to peak values show hints of variation, ($1.30\pm0.50\%$) in
the {\sl I} band, but the detection is not conclusive.
The short data set (1.6 spin periods) means that the uncertainties in this
observation are large. This, coupled with the non-zero peak to peak
values and a tentative detection in the {\sl I} band may warrant
further investigation.

\subsection{DQ Her}
This is the first simultaneous {\sl UBVRI} polarimetric observation of DQ~Her. The
maximum level of circular polarization seen here ($0.64\pm0.31\%$) is
consistent with the plots of \citet{swedlund74} who illustrate a
variation with a max/min of $\gtrsim 0.5\%$. They
found that the polarization was also variable on the orbital period,
our data was just under 0.6 of a complete orbital period so we
cannot bin our data as they did, and we see no overall trend in our
data. Their pass band was approximately equal to our {\sl UBV} bands combined.

The only other measurement of circular polarization in DQ~Her was that
of \citet{stockman92}. They give a broadband integrated result
close to zero, this is likely consistent with \citet{swedlund74} who
see both positive and negative polarization values. As such, this is
the first published result of time-resolved {\sl R} and {\sl I} band (as well as
the first simultaneous {\sl UBVRI}) data. The largest departure from zero
polarization is seen in the {\sl I} band here, and the raw data show up to
$6\pm1\%$.

Although by itself this data cannot claim a significant circular
polarization detection, when considered with the results of
\citet{swedlund74}, it seems likely that
DQ~Her does exhibit variable circular polarization. 
To make a definite conclusion, more measurements are needed, 
particularly in the {\sl I} band.

\subsection{FO Aqr}
This is the first simultaneous {\sl UBVRI} polarimetry of FO~Aqr. All
previous measurements report a mean value close to zero, except for in
the $1.15-1.35\mu m$ range where $+1.1\pm0.3\%$ polarization has been
detected (see Table~\ref{already_known_CP_IPs}). The mean circular polarization 
seen here is within two sigma of all the previous
measurements where there is an overlap in pass band (see
Table~\ref{already_known_CP_IPs}). Since the large value of the
circular polarization in the {\sl I} band has such a large uncertainty we cannot
claim this as a detection, although it is possible that circular polarization
is present at the level of around 1\%.

The short data set (1.1 spin periods) means that the uncertainties are
probably much higher than quoted. This system also perhaps
warrants further investigation, particularly in the {\sl I} band.

\subsection{RXJ1730}
RXJ1730's short spin period (128.0~s) and long orbital period (15.4~hr)
make it a close sibling to the enigmatic AE~Aqr (spin and orbital
periods of 33.1~s and 9.88~h respectively).

The photometric data shows a double peaked profile with equal maxima
and unequal minima (see Fig.~\ref{rxj1730_cp_fot}). Period analysis
yields a spin period of 128.1$\pm$0.7 with the first harmonic
visible at 64.0$\pm$0.2~s. This is in good agreement with \cite{gansicke05} who
concluded that both poles could be seen.

\citet{demartino08} report simultaneous optical Sloan filter data from
the {\sl u'}, {\sl g'} and {\sl r'} bands. The {\sl u'} filter is approximately the same as our
{\sl U} band, {\sl g'} covers
all our {\sl B} and the upper half of {\sl V}, and {\sl r'} covers the lower
half of {\sl V}
as well as {\sl R}. In each of their {\sl u'}, {\sl g'} and {\sl r'} band observations the fundamental and
the first harmonic were seen, with the first harmonic, on average, being
strongest. This is in contrast to what we see (see
Fig.~\ref{rx1730_ubvri_periodograms}), i.e. the fundamental
being dominant. However, \citet{demartino08} show their r' band power spectra
obtained on each of six consecutive nights. This shows a marked change
in the relative strengths of the first harmonic and fundamental over
time, the last night exhibiting a similar structure to
ours. \citet{demartino08} see the strongest signal in their {\sl g'}
and {\sl u'}
bands, our {\sl U} band has very little power, but our {\sl V} band is the
strongest, and since this contributes to what is their {\sl g'} band this
tallies up.

The level of polarization, $8.26\pm1.56$\% peak to peak in the
  {\sl B} band, makes this one of the most variable circularly polarized IPs, and therfore 
likely one of the most magnetic, measured to date. The variation of the circular polarization
in the {\sl B} band, showing both positive and negative values
is indicative of both magnetic poles being visible (since each pole
may only emit either positive or negative circular polarization). We note that the raw circular polarization data is somewhat noisy,
with individual measurements of over 15\%, we are unsure of the origin
of these values, but we speculate that they may arise from short
epochs when the diluting light is randomly lower due to flickering.

In the {\sl B} band the photometry and circular polarization are
  coincident; the peaks in the photometry align with the peak and
  trough in the circular polarization. This strengthens the assertion
  that both poles are seen and they are both emitting circular polarization.

Like many of the other IPs for which circular polarization
has been detected, RXJ1730 is also an {\sl INTEGRAL} source \citep{barlow06}.
We discuss this further in Section 5.8.

The variable nature of this object over the course of several days
\citep{demartino08} makes this an ideal target for a long base line
follow up. Monitoring how the circular polarization varies as the
accretion column structure changes over time and linking this to the
photometry may reveal more about the magnetic nature of this source
and IPs in general.  Phase resolved circular spectro-polarimetry would
be the ideal tool in revealing the magnetic field strength of RXJ1730, but
taking into account the extremely short spin period (128 s), and
relative faintness (V $\sim$17)
there are very few telescope and instrument combinations available where
these kind of observations are possible.

\subsection{V1223 Sgr}
The mean circular polarization in V1223~Sgr is within four sigma of
the previously reported values, and zero (see
Tables~\ref{already_known_CP_IPs} \& \ref{results_summary}). The maximum
peak to peak variation of $2.16\pm1.22\%$ in the {\sl U} band is not
constrained enough for this to be considered a definite
detection, but when considered with the results of \citet{watts85} it is
likely that V1223~Sgr is strongly polarized.

There is a clear hint of a double peaked structure
in the {\sl UVRI} circular polarization curves, indicating that
two magnetic poles can be seen. This is an effect seen in the
photometry also, with hints of a double peaked profile in each
band. This is in contrast to previous results which show a single
peaked structure.

The raw circular polarization measurements are stable in all bands at
the start of the run, but the {\sl U} band begins to fluctuate as the run
goes on, sometimes, quite randomly, up to 20\%.

V1223~Sgr has a similar spin and orbital period to V405~Aur which was 
recently found to be very magnetic \citep{piirola08} and has its
circular polarization peak in the blue part of the spectrum. If the
level of polarization seen here is confirmed then V1223~Sgr would be a
close twin of V405~Aur. We also note that V1223~Sgr is an {\sl INTEGRAL} 
source \citep{barlow06}.

\subsection{V2306 Cyg}

\object{V2306~Cyg} has had significant levels of circular polarization reported previously;
\citet{uslenghi01} first reported it after using the TurPol instrument
at the NOT (results summarised in Table~\ref{already_known_CP_IPs}).
The mean results reported here are within the bounds given by uncertainties
(two sigma) in their {\sl UBV} bands, however in {\sl R} and {\sl I} \citet{uslenghi01}
find a much higher amplitude mean value. \citet{norton02} also
reported {\sl B} and {\sl R} band polarization at the NOT, this time using the
ALFOSC instrument, they obtained mean values of $+0.32 \pm 0.10\%$ and
$-1.99 \pm 0.11\%$ in each band respectively. Here we see significantly
lower values than theirs also.

The shape of the circular polarization variation seen here is one of two minima
per spin cycle in the {\sl B} and {\sl I} bands (see
Fig.~\ref{v2306_cp_fot}). The {\sl B} band of \citet{norton02} has an
indication of a two peaked profile, our results confirm this.

The discrepancy between our results and those reported 
previously can be explained
in a variety of ways. The orbital phase may be different during each
of the observations, \citet{uslenghi01} showed the circular
polarization varied significantly over the orbital period,
unfortunately the orbital ephemeris has accumulated too much uncertainty for
this to be calculated. Another consideration is the brightness of the
source, \citet{uslenghi01} did not quote a magnitude, but
\citet{norton02} measured {\sl UBVRI} magnitudes which are significantly
fainter than ours. Since circular polarization is calculated as a
fraction of the total incoming light this may have had the effect of
seriously diluting our result.

\subsection{Implications of our results}

Generally each of the definite or potential circular polarization detections 
reported here have been most prominent towards the red end of the spectrum. This is 
in agreement with what has been seen before, and reinforces the notion of 
IPs being less magnetic than polars, as weaker fields will give rise to 
polarization appearing at longer wavelengths.

Amongst the IPs detected by {\sl INTEGRAL} \citep{barlow06, bird07} 
RXJ2133 and \object{V2400~Oph} have previously been found to display a large 
degree of circular polarization \citep{katajainen07, buckley95, buckley97}, whilst 
RXJ1730, V2306 Cyg, DQ Her, V1223 Sgr and FO~Aqr are shown here to have some 
degree of circular polarization or at least strong hints of it. The only other 
{\sl INTEGRAL}-detected IP to have its circular polarization measured is
\object{GK~Per} \citep{stockman92}. This was reported as having a
mean value of $0.03\pm0.10\%$, but as noted earlier, the practice of
reporting mean values may seriously under report the true magnetic
nature. In light of this, it would be productive to look 
for circular polarization in the rest of the
{\sl INTEGRAL} IP sources, namely \object{V709~Cas},
\object{IGR000234+6141}, \object{NY~Lup}, and \object{MU~Cam}.
As noted by \citet{katajainen07}, NY~Lup may well be a close
twin of the strongly polarized IP RXJ2133.

It has also been noted that the presence of a soft X-ray component may
be related to the presence of a large magnetic field
\citep{katajainen07}. The circular polarization seen in RXJ1730
further adds to this trend as it is also a soft X-ray source \citep{demartino08}.
This brings the total of soft X-ray emitting, circularly polarized IPs 
to four, namely PQ~Gem, V405~Aur, RXJ2133 and RXJ1730. Perhaps the same
geometry which allows the soft X-ray component to be seen in some
IPs but not others, as suggested by \citet{evans07}, may also allow
the efficient detection of circular polarization. Indeed \citet{evans07} 
suggested that the reason some IPs show polarization and the others
do not, is mostly due to different accretion geometry and hiding effects of
the accretion curtains. This may tie in with the suggestion
by \citet{norton02} with regard to V2306~Cyg, that cancellation of polarized
emission between the two magnetic poles may hide significant polarization
in some systems. Futhermore, it may be that only those systems which show an asymmetry between 
the poles (in terms of temperature etc or accretion curtain structure) or have an offset or non-dipole magnetic
field structure, emit a detectable signal.

\section{Conclusion}

We have detected temporal variation in the circular polarization
emission in RXJ1730, with possible emission (and in some cases variation)
in V2306~Cyg, DQ~Her, V1223~Sgr, AO~Psc and FO~Aqr; AE~Aqr had none
detected at a significant level. 
Broadly speaking this is in agreement with previous results, and adds
to the observational trend of IPs having less polarization than
polars; and hence likely smaller effective magnetic field strength.

There are indications of a correlation between the detection of circular polarization
in IPs and their detection as hard X-ray sources by {\sl INTEGRAL}. We therefore
suggest that other {\sl INTEGRAL} sources should be tested for
circular polarization. Where such 
objects also exhibit soft X-ray components (i.e. NY~Lup and MU~Cam), we predict
there is a very good chance of detecting significant circular polarization.

\section{Acknowledgements}

The Nordic Optical Telescope is operated on the island of La Palma jointly
by Denmark, Finland, Iceland, Norway, and Sweden, in the Spanish
Observatorio del Roque de los Muchachos of the Instituto de
Astrofis\'{\i}ca de Canarias. This work has been supported by the
``Societas Scientiarum Fennica - Suomen Tiedeseura'' and its
Magnus Ehrnrooth foundation, and the Academy of Finland (SK).

\bibliographystyle{aa}
\bibliography{ref}

\longtab{1}{
  \begin{longtable}{llllllll}
    \caption{Summary of previously measured circular polarization in IPs.}\\
    \label{already_known_CP_IPs}\\
    \hline\hline
    Name           & Wavelength range$^a$ & Mean             & Min           & Max            & Integration time  & Total time       & Ref$^1$\\
                    & (nm)                 & (\%)             & (\%)          & (\%)           & (s)               &                  &\\
    \hline
    \endfirsthead

    \caption{continued.}\\
    \hline
    Name           & Wavelength range$^a$ & Mean             & Min           & Max            & Integration time  & Total time       & Ref$^1$\\
                   & (nm)                 & (\%)             & (\%)          & (\%)           & (s)               &                  &\\
    \hline
    \endhead

    \hline
    \endfoot

    \hline
    \multicolumn{8}{l}{$^a$ Numbers in parentheses indicate the effective wavelength of the filter.}\\
    \multicolumn{8}{l}{$^b$ Estimated from plots.} \\
    \multicolumn{8}{l}{$^c$ $I$ corresponds to an unspecified time between 0.5--1~mins.}\\
    \multicolumn{8}{l}{$^d$ Constant polarimeter position.}\\
    \endlastfoot

      AE Aqr        & 350--920             & $-0.03\pm0.02$   &               &                & 15                & 140~min          & 1\\ %\cite{cropper86}\\
                    & {\sl I}              & $-0.06\pm0.03$   &               &                & 15                & 150~min          & 1\\ %\cite{cropper86}\\
                    & 350--570             & $-0.06\pm0.02$   &               &                & 5                 & 390~min          & 1\\ %\cite{cropper86}\\
                    & 320--860             & $+0.01\pm0.02$   &               &                & 8$\times I^c$     & $\times$2        & 2\\ %\cite{stockman92}\\
                    & 590--860             & $-0.01\pm0.01$   &               &                & 1$^d$	     & 34~min           & 2\\ %\cite{stockman92}\\
                    & 320--860             & $-0.13\pm0.03$   &               &                & 8$\times I^c$     & $\times$5        & 2\\ %\cite{stockman92}\\
                    & 320--860             & $+0.01\pm0.01$   &               &                & 8$\times I^c$     & $\times$3        & 2\\ %\cite{stockman92}\\
                    & 1150--1350           & $+0.06\pm0.08$   &               &                &                   & 4~min            & 2\\ %\cite{stockman92}\\
                    & 1450--1650           & $-0.80\pm0.60$   &               &                &                   & 7~min            & 2\\ %\cite{stockman92}\\
                    & 500--750             & $+0.07\pm0.02$   &               &                & 5                 & 255~min          & 3\\ %\cite{beskrovnaya96}\\
                    & (550)                & $+0.06\pm0.01$   &               &                & 5                 & 323~min          & 3\\ %\cite{beskrovnaya96}\\
      AO Psc        & 570--920             & $+0.03\pm0.03$   &               &                & 240               & 180~min          & 1\\ %\cite{cropper86}\\
                    & {\sl I}	             &	            &               &	               & 240	     & 180~min          & 1\\ %\cite{cropper86}\\
                    & 320--860             & $-0.05\pm0.06$   &               &                & 8$\times I^c$     & $\times$1        & 2\\ %\cite{stockman92}\\
                    & 660--860             & $+0.00\pm0.07$   &               &                & 8$\times I^c$     & $\times$3        & 2\\ %\cite{stockman92}\\
                    & 320--860             & $+0.03\pm0.03$   &               &                & 8$\times I^c$     & $\times$2        & 2\\ %\cite{stockman92}\\
      BG CMi        & 640--860             & $-0.24\pm0.03$   &	        &                & 2$\times \sim30$  &                  & 4\\ %\cite{penning86}\\
		& 320--860             & $-0.05\pm0.05$   &               &                &                   &                  & 5\\ %\cite{west87}\\
		& 720--860             & $-0.25\pm0.06$   &               &                &                   &                  & 5\\ %\cite{west87}\\
                    & 1110--1380(1250)     & $-1.74\pm0.26$   &               &                &                   &                  & 5\\ %\cite{west87}\\
                    & 1400--1650(1500)     & $-4.24\pm1.78$   &               &                &                   &                  & 5\\ %\cite{west87}\\
      DQ Her        & 370--580             &                  & $-0.6^b$      & $+0.6^b$       & 14.2              & $\sim$1980~min   & 6\\ %\cite{swedlund74}\\
                    & 320--860             & $+0.01\pm0.01$   &               &                &                   & 250~min          & 2\\ %\cite{stockman92}\\
      EX Hya        & 570--920             & $-0.02\pm0.04$   &               &                & 240               & 200~min          & 1\\ %\cite{cropper86}\\
                    & 590--860             & $+0.01\pm0.02$   &               &                & 2$^d$	     & 68~min           & 2\\ %\cite{stockman92}\\
      FO Aqr        & 640--860             & $+0.06\pm0.02$   &               &                &                   &                  & 4\\ %\cite{penning86}\\
                    & 330--920             & $-0.01\pm0.02$   &               &                & 120               & 200~min          & 1\\ %\cite{cropper86}\\
                    & {\sl I}              & $+0.11\pm0.07$   &               &                & 240               & 200~min          & 1\\ %\cite{cropper86}\\
                    & 320--860             & $-0.06\pm0.04$   &               &                & 8$\times I^c$     & $\times$2        & 2\\ %\cite{stockman92}\\
                    & 320--860             & $+0.01\pm0.04$   &               &                & 8$\times I^c$     & $\times$6        & 2\\ %\cite{stockman92}\\
                    & 720--860             & $-0.01\pm0.17$   &               &                & 8$\times I^c$     & $\times$5        & 2\\ %\cite{stockman92}\\
                    & 1150--1350           & $+0.19\pm0.13$   &               &                &                   & 51~min           & 2\\ %\cite{stockman92}\\
                    & 1150--1350           & $+1.09\pm0.31$   &               &                &                   & 141~min          & 2\\ %\cite{stockman92}\\
		& Visual               & $+0.11\pm0.13$   & $-0.2^b$      & $+0.3^b$       & 28                & 270~min          & 7\\ %\cite{berriman86}
		& IR                   & $-0.01\pm0.55$   & $-1.3^b$      & $+0.8^b$       & 28                & 270~min          & 7\\ %\cite{berriman86}
      GK Per        & 1150--1350           & $+0.03\pm0.10$   &               &                &                   & 45~min           & 2\\ %\cite{stockman92}\\
      PQ Gem        & {\sl U}              & $+0.0\pm0.6$     &               &                &                   & 80~min           & 8\\ %\cite{rosen93}\\
                    & {\sl B}              & $+0.0\pm0.6$     &               &                &                   & 80~min           & 8\\ %\cite{rosen93}\\
                    & {\sl V}              & $+0.0\pm0.9$     &               &                &                   & 80~min           & 8\\ %\cite{rosen93}\\
                    & {\sl R}              &                  & $-1.1$        & $+0.6$         &                   & 80~min           & 8\\ %\cite{rosen93}\\
                    & {\sl U}              &                  & $-0.4^b$      & $+0.3^b$       & 8$\times 5$       & 162~min          & 9\\ %\cite{piirola93}\\
                    & {\sl B}              &                  & $-0.3^b$      & $+0.4^b$       & 8$\times 5$       & 162~min          & 9\\ %\cite{piirola93}\\
                    & {\sl V}              &                  & $-0.7^b$      & $+0.7^b$       & 8$\times 5$       & 162~min          & 9\\ %\cite{piirola93}\\
                    & {\sl R}              &                  & $-1.5^b$      & $+0.7^b$       & 8$\times 5$       & 162~min          & 9\\ %\cite{piirola93}\\
                    & {\sl I}              &                  & $-2.7$        & $+1.5$         & 8$\times 5$       & 162~min          & 9\\ %\cite{piirola93}\\
                    & {\sl U}              &                  & $-0.5^b$      & $+0.3^b$       &                   & $\sim$730~min    & 10\\ %\cite{potter97}\\
                    & {\sl B}              &                  & $-0.2^b$      & $+0.3^b$       &                   & $\sim$730~min    & 10\\ %\cite{potter97}\\
                    & {\sl V}              &                  & $-0.6^b$      & $+0.4^b$       &                   & $\sim$730~min    & 10\\ %\cite{potter97}\\
                    & {\sl R}              &                  & $-1.0^b$      & $+0.2^b$       &                   & $\sim$730~min    & 10\\ %\cite{potter97}\\
                    & {\sl I}              &                  & $-1.3^b$      & $+1.3^b$       &                   & $\sim$730~min    & 10\\ %\cite{potter97}\\
                    & {\sl J}              &                  & $-1.2^b$      & $+1.0^b$       &                   & $\sim$270~min    & 10\\ %\cite{potter97}\\
                    & {\sl K}              &                  & $-2.0^b$      & $+1.3^b$       &                   & $\sim$460~min    & 10\\ %\cite{potter97}\\
      RXJ2133       & (360)                & $+0.90\pm0.06$   & $-0.2^b$      & $+1.5$         & 4$\times$24       & 229~min          & 11\\ %\cite{katajainen07}\\
                    & (440)                & $+1.12\pm0.05$   & $+0.2^b$      & $+2.5$         & 4$\times$24       & 229~min          & 11\\ %\cite{katajainen07}\\
                    & (530)                & $+1.17\pm0.09$   & $-0.3^b$      & $+3.5$         & 4$\times$24       & 229~min          & 11\\ %\cite{katajainen07}\\
                    & (690)                & $+0.85\pm0.07$   & $+0.2^b$      & $+3$           & 4$\times$24       & 229~min          & 11\\ %\cite{katajainen07}\\
                    & (830)                & $+0.89\pm0.08$   & $-0.4^b$      & $+2.5$         & 4$\times$24       & 229~min          & 11\\ %\cite{katajainen07}\\
      TV Col        & 640--860             & $-0.03\pm0.04$   &               &                &                   &                  & 4\\ %\cite{penning86}\\
                    & 320--860             & $-0.13\pm0.09$   &               &                & 8$\times I^c$     & $\times$1        & 2\\ %\cite{stockman92}\\
                    & 320--860             & $-0.07\pm0.07$   &               &                & 8$\times I^c$     & $\times$3        & 2\\ %\cite{stockman92}\\
                    & 320--860             & $-0.08\pm0.10$   &               &                & 8$\times I^c$     & $\times$1        & 2\\ %\cite{stockman92}\\
      V1223 Sgr     & 350--920             & $-0.06\pm0.03$   &               &                & 240               & 144~min          & 1\\ %\cite{cropper86}\\
                    & {\sl I}+ {\sl R}     & $-0.04\pm0.07$   &               &                & 240               & 563~min          & 1\\ %\cite{cropper86}\\
	          & {\sl V}	             & $-0.48\pm0.62$   & $\gtrsim-2$   & $\lesssim+2$   & 14	               &		    & 12\\ %\cite{watts85}
	          & {\sl R}	             & $+0.03\pm0.13$   & $\gtrsim-0.5$ & $\lesssim+0.5$ & 14	               &	              & 12\\ %\cite{watts85}
	          & {\sl J}	             & $-0.36\pm0.13$   & $\gtrsim-1$   & $\lesssim+0.5$ & 14	               &		    & 12\\ %\cite{watts85}
	          & {\sl K}	             & $+1.14\pm1.26$   & $\gtrsim-8$   & $\lesssim+8$   & 14	               &		    & 12\\ %\cite{watts85}
      V2306 Cyg     & (360)                & $+0.04\pm0.06$   &               &                & 8$\times$10 (210) & 872~min          & 13\\ %\cite{uslenghi01}\\
                    & (440)                & $+0.16\pm0.08$   &               &                & 8$\times$10 (210) & 872~min          & 13\\ %\cite{uslenghi01}\\
                    & (530)                & $+0.18\pm0.11$   &               &                & 8$\times$10 (210) & 872~min          & 13\\ %\cite{uslenghi01}\\
                    & (690)                & $-0.55\pm0.08$   & $-1.3^b$      & $+1^b$         & 8$\times$10 (210) & 872~min          & 13\\ %\cite{uslenghi01}\\
                    & (830)                & $-0.91\pm0.14$   & $-1.7^b$      & $+0.4^b$       & 8$\times$10 (210) & 872~min          & 13\\ %\cite{uslenghi01}\\
                    & {\sl B}              & $+0.32\pm0.10$   & $-0.3^b$      & $+0.7^b$       & 58                & 52~min           & 14\\ %\cite{norton02}\\
                    & {\sl R}              & $-1.99\pm0.11$   & $-5.2^b$      & $-0.6^b$       & 45                & 55~min           & 14\\ %\cite{norton02}\\
      V2400 Oph     & WL                   &                  & $-2.9^b$      & $-1.0^b$       & 120 \& 180        & $\sim$900~min    & 15\\ %\cite{buckley95}\\
                    & {\sl V}              & $\sim-1.8^b$     & $-4.8^b$      & $-1.0^b$       & 120               & $\sim$33~min     & 15\\ %\cite{buckley95}\\
                    & {\sl R}              & $\sim-2.3^b$     & $-5.1^b$      & $-0.5^b$       & 120               & $\sim$71~min     & 15\\ %\cite{buckley95}\\
                    & {\sl I}              & $\sim-3.3^b$     & $-6.0^b$      & $-1.0^b$       & 120               & $\sim$32~min     & 15\\ %\cite{buckley95}\\
                    & 320--700 (470)       & $-0.90\pm0.03$   &               &                & 50                &                  & 16\\ %\cite{buckley97}\\
                    & 560--900 (700)       & $-2.82\pm0.04$   &               &                & 50                &                  & 16\\ %\cite{buckley97}\\
      V405 Aur      & 500--750             & 1.8 (Semi-amp)   &               &                &                   &                  & 17\\ %\cite{shakhovskoj97}\\
                    & {\sl U}              &                  & $-2$          & $+2$           & 8$\times$12 (96)  & 1328~min         & 18\\ %\cite{piirola08}
                    & {\sl B}              &                  & $-3$          & $+3$           & 8$\times$12 (96)  & 1328~min         & 18\\ %\cite{piirola08}
                    & {\sl V}              &                  & $-3$          & $+3$           & 8$\times$12 (96)  & 1328~min         & 18\\ %\cite{piirola08}
                    & {\sl R}              &                  & $-2$          & $+2$           & 8$\times$12 (96)  & 1328~min         & 18\\ %\cite{piirola08}
                    & {\sl I}              &                  & $-1$          & $+1$           & 8$\times$12 (96)  & 1328~min         & 18\\ %\cite{piirola08}
      YY Dra        & 320--860             & $+0.09\pm0.10$   &               &                & 8$\times I^c$     & $\times$1        & 2\\  %\cite{stockman92}
      \end{longtable}

\footnotetext[1]{References - (1) \citet{cropper86}; (2) \citet{stockman92}; (3) \citet{beskrovnaya96}; (4) \citet{penning86}; (5) \citet{west87}; (6) \citet{swedlund74}; (7) \citet{berriman86}; (8) \citet{rosen93}; (9) \citet{piirola93}; (10) \citet{potter97}; (11) \citet{katajainen07}; (12) \citet{watts85}; (13) \citet{uslenghi01}; (14) \citet{norton02}; (15) \citet{buckley95}; (16) \citet{buckley97}; (17) \citet{shakhovskoj97}; (18) \citet{piirola08}.}
}
\end{document}